\documentclass[a4paper,11pt]{article}
\usepackage{jcappub} 
\usepackage{lineno}
\usepackage{tikz}
\usepackage[english]{babel}
\usepackage[utf8]{inputenc}
\usepackage[colorinlistoftodos, color=green!40, prependcaption]{todonotes}
\usepackage{appendix}
\usepackage{amssymb,mathtools}
\usepackage{subfigure}
\usepackage{soul}
\usepackage[normalem]{ulem}

\arxivnumber{2411.11439} 
\title{\boldmath Interpreting the LHAASO Galactic diffuse emission data}

\author[a,1]{V. Vecchiotti\note{Corresponding author.}}
\author[b]{and G. Peron}
\author[b]{and E. Amato}
\author[c]{and S. Menchiari}
\author[b]{and G. Morlino}
\author[d]{and G. Pagliaroli}
\author[d,e]{and F.L. Villante}

\affiliation[a]{Department of Physics, NTNU,\\
NO-7491 Trondheim, Norway}
\affiliation[b]{INAF Osservatorio Astrofisico di Arcetri Largo Enrico Fermi,5, 50125, Firenze, Italy}
\affiliation[c]{Instituto de Astrofísica de Andalucía, CSIC, 18080 Granada, Spain}
\affiliation[d]{INFN, Laboratori Nazionali del Gran Sasso,\\
67100 Assergi (AQ),  Italy}
\affiliation[e]{Department of Physical and Chemical Sciences, University of L'Aquila, \\
67100 L'Aquila, Italy}

\emailAdd{vittoria.vecchiotti@ntnu.no}

\abstract{Recently, the Large High-Altitude Air Shower Observatory (LHAASO) collaboration has obtained a measurement of the gamma-ray diffuse emission in the ultra-high energy range, $10-10^3$ TeV after masking the contribution of known sources. 
The measurement is provided in two Galactic regions and appears to be 2–3 times higher than the gamma-ray signal expected from the hadronic interactions of diffuse cosmic rays with the interstellar medium, potentially implying that either additional emission sources exist or cosmic ray intensities have spatial variations.
In this work, we calculate the hadronic gamma-ray diffuse emission outside the masks, considering a realistic gas distribution. We present a comprehensive calculation of the emission, which includes systematic uncertainties in the gas content of the Galactic disk, in the energy and spatial distribution of cosmic rays, as well as in the hadronic interaction cross-sections.
Our results show that these factors mitigate the tension between data and predictions.
The LHAASO data appear compatible with our baseline model in the outer Galactic region.
In the inner region, the data show an excess with respect to the predictions below $\sim 50$ TeV, while at higher energies they are well described by our model.
We argue that two plausible explanations for enhanced gamma-ray emission—unresolved sources and CR spectral hardening in the inner Galaxy—are likely suppressed by the LHAASO masking strategy, which excludes regions where both effects are expected to be most prominent.}

\begin{document}
\maketitle
\flushbottom

\section{Introduction}

The study of gamma-ray (and neutrino) emission produced by cosmic rays (CRs) interacting with the interstellar medium (ISM) after escaping the accelerator is of fundamental importance to constrain the CR origin and propagation.
In the last years, there have been an increasing number of attempts to measure this \textquotedblleft truly diffuse'' component at very high energy, both by Imaging Atmospheric Cherenkov Telescopes, like H.E.S.S. \cite{HESS:2014ree}, and air shower arrays, such as ARGO-YBJ \cite{ARGO-YBJ:2015cpa}, Milagro \cite{Milagro2008ApJ} and HAWC \cite{HAWC:2023wdq}.
This has allowed us to extend our knowledge up to the sub-PeV energy domain, recently probed by the Tibet AS$\gamma$ \cite{TibetASgamma:2021tpz} and LHAASO \cite{LHAASO:2023diff} detectors.
However, a recurrent caveat with the experimental determination of the diffuse component relates to the subtraction of individual sources from the observed emission: this is bound to leave contamination by unresolved sources, too faint or too extended to be detected.
%
At high energy, this component is naturally expected to be non-negligible and it must be included in the theoretical modeling of the data \cite{Cataldo:2020qla, Steppa:2020qwe, Vecchiotti:2021vxp,Vecchiotti:Tibet}. 

Such an effect has been already discussed in the context of the data released by the Tibet AS$\gamma$ experiment, which provided the first measurement of the diffuse gamma-ray flux between $10^2-10^3$ TeV in two different regions of the Galactic plane \cite{TibetASgamma:2021tpz}.
In that work, the diffuse emission was obtained by subtracting only the flux produced by known TeV sources, as listed in the TeVCat catalog \cite{2008ICRC....3.1341W}.
However, it was noticed that a non-negligible contribution from an unresolved population of TeV sources is expected to contaminate the observations in both regions \cite{Fang:2021ylv, Vecchiotti:Tibet}. Indeed, when such a contribution is included, the agreement between the theoretical models and the observed data improves considerably.
%
Alternatively, 
one may speculate that CR diffuse emission is larger than standard expectations because of a progressive hardening of the CR spectrum toward the Galactic center, a possibility recently emerged from the analysis of {\it Fermi}-LAT data \cite{Acero:2016qlg, Yang:2016jda, Pothast:2018bvh},  which challenges the idea that CR diffusion is spatially uniform in the Galactic disk\footnote{
It is to be emphasized that the CR spectral hardening hypothesis, differently from the unresolved source component, can only improve theoretical predictions in the inner Galactic region probed by Tibet AS$\gamma$ \cite{Vecchiotti:Tibet}, while it would not affect the emission from the external part of the Galaxy.}.

Recently, the LHAASO collaboration reported a measurement of the diffuse Galactic gamma-ray emission in the energy range $10-10^3$ TeV in two different regions of the sky using the data recorded by the square kilometer array (KM2A) \cite{LHAASO:2023diff}.
These data are obtained by LHAASO using a different procedure with respect to Tibet AS$\gamma$, namely by masking (rather than subtracting) the known TeV sources.
This includes all objects listed in TeVCat as well as the new sources detected at ultra and very high energy by the KM2A detector of the LHAASO experiment \cite{LHAASO:2023cat}.
The flux in the non-masked regions was then compared by the LHAASO collaboration with theoretical predictions for CR diffuse emission, finding that measurements exceed expectations by a factor of 3 (2) in the inner (outer) region of the Galaxy.
Then, in a follow-up study, \cite{Zhang:2023ajh} argued that the theoretical predictions based on the locally observed CR spectrum are not able to explain the combined observations of {\it Fermi}-LAT and LHAASO from several GeV and up to $\sim 60$ TeV.
It was suggested that the observed excess might result from unresolved sources \cite{Shao:2023aoi}, possibly associated with Pulsar Wind Nebulae (PWNe) and TeV halos \cite{Yan:2023hpt, Dekker:2023six} or star clusters \cite{Menchiari:2024uce}. 

However, to correctly assess the role of unresolved sources, as well as possible spatial variations of the CR spectrum throughout the Galactic plane, a careful evaluation of the uncertainties involved in the estimate of the diffuse emission must be carried out. 
This is precisely the aim on this paper, in which we focus on modeling the total diffuse emission (namely ``truly diffuse'' plus unresolved sources) by accounting for uncertainties in the $pp$ inelastic cross-section, the gas distribution in the Galactic plane, and the CR spectrum detected at Earth at energies $\gtrsim 100$\, TeV, where the measurements provided by different experiments give different spectra for protons.

We additionally test the possibility that the spatial dependence of the CR spectral index, inferred from the analysis of the large-scale diffuse emission measured by {\it Fermi}-LAT at $\sim 20$\,GeV \cite{Acero:2016qlg, Yang:2016jda, Pothast:2018bvh}, extends to the energy range, $>10$ TeV, probed by LHAASO.

We then consider the possible contribution of unresolved sources to the LHAASO measurement. To this aim, we consider the source population by \cite{Cataldo:2020qla}, based on the TeV sources included in the H.E.S.S. Galactic Plane Survey (H.G.P.S.) \cite{H.E.S.S.:2018zkf}.
We here show that the same source population reproduces, within 2$\sigma$, the differential flux at $50$ TeV of the sources listed in the KM2A catalog, and assess its overall contribution to the detected large-scale emission.

The plan of the paper is as follows. In Sec.~\ref{Sec:diffuse}, we introduce our model to estimate the Galactic diffuse gamma-ray emission and the different assumptions on the $pp$ inelastic cross-section, the ISM, and the CR spectrum. In Sec.~\ref{Sec:Unresolved}, we describe our synthetic  population and compare the results of our model with LHAASO data, both in terms of number of sources as well as their total flux.
In Sec.~\ref{sec:Results}, we show our results. In particular, in Sec.~\ref{sec:results diffuse} we compare our diffuse emission models with the LHAASO diffuse emission data.
In Sec.~\ref{Sec:hardening} we assess the impact of a CR spectral index that depends on the Galactocentric radius. In Sec.~\ref{sec:unresolved}, we assess the contribution of unresolved sources. Finally, in Sec.~\ref{sec:Conclusions} we draw our conclusions.

\section{Diffuse emission}
\label{Sec:diffuse}
The interaction with the ISM of accelerated protons and heavier nuclei propagated in the Galactic magnetic field results in high energy gamma-ray and neutrino diffuse emission.
Providing a theoretical estimate of this component is of fundamental importance to interpret the data and requires a careful treatment of all the involved uncertainties.

The gamma-ray diffuse emission as a function of the photon energy, $E_{\gamma}$, and arrival direction, $\hat{n}_{\gamma}$, is parametrized in the following way \cite{Pagliaroli:2016, Cataldo:2019qnz}:
\begin{eqnarray} \label{gamma diffuse flux}
\nonumber
\varphi_{\gamma,{\rm diff}}&&(E_{\gamma},\hat{n}_{\gamma}) =   \int_{E_{\gamma}}^{\infty} dE\; \frac{d \sigma(E,E_{ \gamma})}{dE_{\gamma}} \times \\  
&&
\hspace{-1.5cm}
\int_{0}^{\infty} dl\; \varphi_{\rm CR}(E ,{\bf r_{\odot}} + l \hat{n}_{\gamma})\, n_{\rm H}({\bf r_{\odot}} + l\hat{n}_{\gamma})e^{-\tau(E_{\gamma}, l \hat{n}_{\gamma})} ,  
\end{eqnarray}
where $\frac{d \sigma(E, E_{\gamma})}{dE_{\gamma}}$ is the differential cross section for the photon production as a function of the CR energy $E$; $\varphi_{\rm CR}(E ,{\bf r})$ is the CR flux as a function of energy and position in the Galaxy, ${\bf r}$, where ${\bf r} =  {\bf r_{\odot}} + l \hat{n}_{\gamma}$ with ${\bf r_{\odot}}=8.5$ kpc the position of the Sun with respect to the Galactic center and $l$ the length of the line of sight.
Finally, $n_{\rm H}({\bf r})$ is the number density of target nucleons, while the exponential function takes into account the photon absorption in the interstellar radiation field (mainly due to the Cosmic Microwave Background radiation) which suppresses the flux produced at large distances \cite{Vernetto:2016alq}. This process suppresses mainly the gamma-ray emission at energies larger than $E_{\gamma} \ge 100\; {\rm TeV}$.
The integral is performed over the nucleon energy $E$ and along the line of sight $l$.


The modeling of the diffuse emission is subject to significant uncertainties, related to the quantities listed above, that need to be taken into account \cite{Schwefer:2022zly}.
In this work, we adopt a fiducial model for the diffuse gamma-ray emission (see Sec.~\ref{sec: fiducial case}) and we estimate the variations induced by theoretical and observational uncertainties with respect to this, as described in the following subsections.

\subsection{Cross-section}
Concerning the $pp$ cross-section, following the same approach as \cite{Schwefer:2022zly}, we estimate the uncertainty by considering different parametrizations: the parametrizations provided by \cite{Kafexhiu:2014cua} based on the Monte Carlo code Pythia 8.18,
and the most recent parametrization, by \cite{Kachelriess:2022khq}, through the AAFRAG interpolation routines based on QGSJET-II-04m. A detailed comparison between these cross-sections and other parameterizations available in the literature can be found in Appendix \ref{App:cross-sec}.

Above a few TeV, which is the energy of interest for this work, the diffuse gamma-ray emission is maximized when using the Pythia 8.18 parametrization \cite{Kafexhiu:2014cua}, while the minimum is obtained by using the AAFRAG parametrization \cite{Kachelriess:2022khq} (see Appendix \ref{App:cross-sec}).

\subsection{Gas distribution}
\label{sec:gas}
To estimate the uncertainty due to the gas distribution, we use two different models as derived from different gas tracers. The first one is taken from the ancillary data provided along with the GALPROP  code \cite{Galprop} (we hereafter refer to it as GALPROP maps) and includes the contributions from atomic, $\rm{H}$, and molecular hydrogen, $\rm{H}_{2}$, traced by the ${\rm H}$\textsc{i,} \cite{Kalberla2005,hi4pi2016} and the CO emission line \cite{Dame2001}, respectively. A uniform $X_{CO}$ conversion factor $1.9\times 10^{20}\, \rm cm^{-2}\, K^{-1}\, km^{-1}\, s$ is assumed between CO brightness and H$_2$  column density, while the HI is calibrated assuming a uniform spin temperature of $125$ K \cite{Strong2004}.
The second model is based on the hydrogen gas column density ($N_H$) traced by the PLANCK dust opacity ($\tau_D$) map obtained at 353 GHz \footnote{\textit{Based on observations obtained with Planck (\href{http://www.esa.int/Planck}{http://www.esa.int/Planck}), an ESA science mission with instruments and contributions directly funded by ESA Member States, NASA, and Canada.}}\cite{Planck:2016frx}. 
The latter can be used as a gas tracer, since dust is uniformly mixed with neutral gas. 
The dust-to-gas conversion factor is calibrated on experimental data, and, for our case, we use   $X^{-1}_D \equiv \big({\tau_D}/{N_H}\big)= 1.18 \times 10^{-26} \rm{cm^{2}}$, as reported by \cite{Planck2011}.
In order to account for the presence of heavier elements, we multiply the hydrogen density by a factor 1.42, which reflects the Solar System composition, here assumed to be representative of the entire Galactic Disk \cite{Ferriere:2001rg}. 

The Galactic diffuse emission depends on the three-dimensional distribution of the gas, as we consider a position-dependent CR distribution and account for the $\gamma\gamma$ absorption of high-energy photons on the interstellar radiation fields. 
For the first model, we derive the three-dimensional spatial distribution of gas directly from the GALPROP maps, which we use as a reference in what follows. The maps provide a 3D decomposition in terms of longitude, latitude, and distance, obtained from the radial velocity measurements contained in the original CO and HI maps. 
Conversely, being the dust map a measure of opacity, it traces the column density integrated along the line of sight and does not provide any information about the 3D gas distribution. 
Hence, for the second model, we adopt the same spatial dependence as the GALPROP maps. 
As a consequence, the two models differ only by a normalization factor, which, however, varies for different lines of sight.

When integrated along the line of sight, in the regions considered by LHAASO, the total column density obtained with the GALPROP maps is, on average, about $20\%$ ($50\%$) larger than the column density derived from the dust maps in the inner (outer) region.
    
\subsection{Cosmic rays}
The CR flux is parameterized as the product of three terms:
\begin{equation}
\varphi_{\rm CR}(E,{\bf r}) = \varphi_{\rm CR,\odot}(E)\,g({\bf r})\,h({E,\bf r})\ ,
\label{Eq:CR_flux}
\end{equation}
where $\varphi_{\rm CR,\odot}(E)$ represents the flux of nucleons measured at the Earth, $g({\bf r})$ describes the spatial distribution of CRs through the Galaxy.  
The function $g({\bf r})$ is dimensionless and normalized to 1 at the Sun position ${\bf r}_\odot$.
The function $h({E,\bf r})$ introduces a position-dependent variation of the CR spectral index, aimed at including the possible spectral hardening of large-scale gamma-ray emission in the inner Galaxy that was earlier inferred from analysis of the {\it Fermi}-LAT data \cite{Acero:2016qlg, Yang:2016jda, Pothast:2018bvh}.

For the local CR nucleon flux, $\varphi_{\rm CR,\odot}(E)$, we consider the data-driven parameterization provided by \cite{Dembinski:2017}, which includes the contribution from all the nuclear species. 
These data are also affected by uncertainties. Indeed the CR proton spectra measured by IceTop \cite{IceCube:2019hmk} and KASKADE \cite{Apel:2013uni} above $1$ PeV are not in agreement. The fit performed by \cite{Dembinski:2017} to CR protons well reproduces the IceTop data, but overpredicts the KASKADE data above $1$ PeV.
In order to take into account the observational uncertainty in the proton spectrum, we consider an additional case in which the proton best fit from \cite{Dembinski:2017} is replaced by a fit that reproduces the KASKADE data, while the fits for heavier nuclei remain unchanged (see Appendix~\ref{App:Fit} for details).
%
We anticipate that the difference between the two spectra is such that the gamma-ray diffuse emission varies by a factor of $\sim 2$ at $500$ TeV.

An additional source of uncertainty is the spatial distribution of CRs in our Galaxy, $g({\bf r})$.
In this work, we consider two cases: 1) $g({\bf r})=1$, namely the simplest scenario, in which CRs are uniformly distributed in the Galaxy; 2) $g({\bf r}) = g_{\rm snr}({\bf r})$ defined as the solution of a 3D isotropic diffusion equation with constant diffusion length, $R$, and stationary CR injection, proportional to the source distribution, $f_{\rm S}({\bf r})$. In this latter case, we adopt the same parametrization as \cite{Cataldo:2019qnz}, with $f_{\rm S}({\bf r})$ following the SNR number density provided by \cite{Green:2015isa} and diffusion length $R$ fixed to infinity\footnote{This is the value that better reproduces the gamma-ray emissivity data provided by {\it Fermi}-LAT at $20$ GeV \cite{Cataldo:2019qnz}.}.

%
In the two regions of interest, the total diffuse emission seems to depend only slightly on the different assumptions on $g({\bf r})$. In particular, we find that the flux varies by less than $10\%$ in the inner region and by about $20\%$ in the outer region, as a consequence of these assumptions on $g({\bf r})$.

As the last ingredient, the function $h({E,\bf r})$ introduces the possibility that the CR spectral index is position-dependent. 
The function is defined as:
\begin{equation}
h(E,{\bf r})=\left(\frac{E}{\overline{E}}\right)^{\Delta({\bf r})}
\label{Eq:h_funct}
\end{equation}
where $\overline{E}=20\,{\rm GeV}$  is the pivot energy and $\Delta({\bf r}_\odot)=0$. 
The function $\Delta({\bf r})$ in Galactic cylindrical coordinates is modeled as:
\begin{equation}
  \Delta(r,z) =\Delta_0
  \begin{cases}
    \left(1 - \frac{r}{r_{\odot}} \right) & r\le 10\, \rm kpc \\
    \left(1 - \frac{10\, \rm kpc}{r_{\odot}} \right) & r > 10\, \rm kpc
  \end{cases}
\label{Eq:Delta} 
\end{equation}
The factor $\Delta_0=0.3$ represents the difference between the CR spectral index at the Galactic center and its value at the Sun position. 
This choice enables us to reproduce the spectral hardening of large-scale gamma-ray emission from the inner Galaxy, inferred from {\it Fermi}-LAT data by \cite{Acero:2016qlg, Yang:2016jda, Pothast:2018bvh}. Our description is essentially equivalent to the approach followed by \cite{Lipari:2018gzn} in their ``space-dependent" model, and in agreement with the results obtained by \cite{Gaggero:2014xla, Gaggero:2015, Gaggero:2017jts} in their KRA$\gamma$ CR propagation model.

\subsection{Fiducial model}
\label{sec: fiducial case}
Our fiducial model is defined using the following hypotheses: 
\begin{enumerate}
    \item Cross-section: AAFRAG \cite{Kachelriess:2022khq};
    \item Gas-template: GALPROP maps \cite{Strong2004};
    \item CR spectrum parametrization: from \cite{Dembinski:2017};
    \item CR spatial distribution: from \cite{Cataldo:2019qnz}, i.e., $g({\bf r}) = g_{\rm snr}({\bf r})$;
    \item Standard diffusion, without hardening: $h(E,{\bf r})=1$.
\end{enumerate}
%
In what follows, different choices of parametrization for items $1-4$, with respect to this reference case, define the theoretical uncertainties. The effect of the hardening is discussed separately.

\section{Galactic PWNe population}
\label{Sec:Unresolved}
In order to calculate the contribution of unresolved sources to the LHAASO diffuse gamma-ray emission, we build a synthetic population of gamma-ray sources. 
Following \cite{Cataldo:2020qla}, we consider only sources powered by pulsar activity whose intrinsic luminosity $L$ (integrated in the $1-100$ TeV energy range) decreases as a function of the source age $t_{\rm age}$ according to:
\begin{equation}
L(t_{\rm age})= L_{\rm max}\left(1+\frac{t_{\rm age}}{\tau_{\rm sd}}\right)^{-\frac{n+1}{n-1}},
\label{lum}
\end{equation}
where $L_{\rm max}$ is the maximum luminosity and $\tau_{\rm sd}$ is the pulsar spin-down timescale.
The parameter $n$ is the braking index. In particular, we fix $n=3$ which corresponds to assuming that pulsars lose their energy via magnetic dipole radiation.
We sample the source age uniformly in the interval $[0,\, T]$, where $T=10^6$~yr is the assumed duration of very high-energy emission, and consider the sources to be spatially distributed according to a density function $\rho({\bf r})$.
The latter is assumed to be proportional to the pulsar distribution parameterized by \cite{Lorimer:2006qs} and to scale as $\exp \left(-\left|z  \right|/H\right)$ along the direction $z$ perpendicular to the Galactic plane, where $H$ represents the thickness of the disk. 
The total number of sources in our population is fixed by ${\cal R} \, T$, where ${\cal R}=0.019\,{\rm yr}^{-1}$ is the rate of core-collapse SNe in our Galaxy \cite{Diehl:2006cf}.
The above approach is physically well-motivated for sources powered by pulsar activity, such as PWNe or TeV halos. 
%
It was shown in \cite{Cataldo:2020qla} that the parameters $L_{\rm max}$ and $\tau_{\rm sd}$ that define the luminosity function of the considered population can be effectively constrained by fitting the longitude, latitude and integrated flux distribution of bright sources included in the H.G.P.S. catalog.
We consider here the values $L_{\rm max}=2.23\times 10^{35}\, \rm erg \, \rm s^{-1}$ and $\tau_{\rm sd}=2.9$ kyr that are obtained by assuming that the thickness of the Galactic disk is equal to $H=0.05\ {\rm kpc}$.
This choice is motivated by the fact that it provides the best $\Delta\chi^{2}$ in the fit of the H.G.P.S catalog, as can be seen in Tab.~1 of \cite{Cataldo:2020qla} \cite[see also][]{Pagliaroli:2023naa}.
Moreover, the above values are obtained by assuming that all sources have power-law spectra with spectral index equal to $2.3$ in the 1-100 TeV energy range~\cite{H.E.S.S.:2018zkf}.
Nonetheless, we can change the spectral assumptions without spoiling the agreement with the H.G.P.S. data.
We are allowed to do so because the best-fit value of $\tau_{\rm sd}$ does not depend on the assumed spectrum, while the best-fit value for $L_{\rm max}$ can be modified in such a way to ensure that the maximum TeV emissivity of the population, $F_{\rm max}$, remains constant.
Specifically, we can write $F_{\rm max} = \frac{L_{\rm max}}{\langle E \rangle}$, where $\langle E \rangle$ denotes the average energy of photons emitted in the range $1-100\,{\rm TeV}$, and adjust $L_{\rm max}$ to compensate for changes in $\langle E \rangle$ resulting from the different spectral assumptions \cite{Vecchiotti:2023nqy}.
%
%
For the spectral assumption adopted by \cite{Cataldo:2020qla}, one has $\langle E \rangle = 3.25$ TeV.
In this work, we assume that all sources have a power-law spectrum with spectral index $2.4$, which is the average spectral index of the brightest sources observed by H.E.S.S., with an exponential cut-off at $E_{\rm cut}=100$ TeV.
The cut-off is introduced because the LHAASO-KM2A measurements extend to higher energies than those of H.E.S.S.
The specific value of the cut-off is justified later in this section.
With this spectral assumption, one obtains $\langle E \rangle = 2.73$ TeV, and correspondingly the maximal luminosity is shifted from the above-quoted value to $L_{\rm max} = 1.87 \times 10^{35} \, \rm erg \, \rm s^{-1}$.

In order to calculate the unresolved source contribution, we consider the KM2A sensitivity threshold for point-like sources and spectral index $3$ as provided by the LHAASO collaboration in Fig.~5 of \cite{LHAASO:2023cat}.
Although the sensitivity threshold is calculated for a specific value of the spectral index, the dependence on the latter is only weak, as can be seen by comparing the sensitivity curves in Fig.~5 of \cite{LHAASO:2023cat} for different spectral assumptions.
Thus, we can safely apply it to our synthetic population.

In our reference case, all sources are considered point-like. 
However, we cannot exclude the possibility that at least some fraction of sources may be extended, such as TeV halos that could be present around some PWNe. 
This is important because it affects the number of unresolved sources whose flux contributes to the diffuse emission. 
To evaluate the impact of a population of extended sources, we consider a second case where all sources have a physical size of $40$ pc that we built by using the appropriate best-fit values of $L_{\rm max}=2.02\times 10^{35}\, \rm erg \, \rm s^{-1}$ and $\tau_{\rm sd}=4.6$ kyr. 
%
In this case, we scale the point-like threshold with the factor $\sqrt{(\sigma_{\rm s}^2+\sigma_{\rm psf}^2)/\sigma_{\rm psf}^2}$, where $\sigma_{\rm s}$ is the angular size of the sources and $\sigma_{\rm psf}=0.2^{\circ}$ is the point spread function of the LHAASO experiment at 50 TeV \cite{LHAASO:2023cat}.
This represents a good approximation of the sensitivity for extended sources in the background-dominated energy range, i.e., below the LHAASO sensitivity minimum at 100 TeV \cite[see discussion in][]{Celli:2024cny}.

\subsection{Compatibility with LHAASO-KM2A}
\label{sec:compatibility}
We build 100 realization of our synthetic population of sources and, for each one, we calculate the number of sources that would have been resolved by LHAASO-KM2A and the total flux from both resolved and unresolved sources at 50~TeV. 
The results are reported in Tab.~\ref{tabLHAASO} and ~\ref{tabLHAASO40} for the cases of point-like and extended sources, respectively.
In particular, the predictions are provided both in the specific regions where diffuse emission measurements are available, i.e., $15^{\circ}<l<125^{\circ}$ (inner region) and $125^{\circ}<l<235^{\circ}$ (outer region) for $|b|<5^{\circ}$, as well as in the entire longitudinal window probed by LHAASO.
The number of resolved sources and the associated errors are obtained as the mean and variance over the 100 Monte Carlo realizations.
The fluxes instead are obtained as the median values over the realizations, with the lower and upper errors representing the 25 and 75 percentile, respectively\footnote{We display the median value and interquartile range, instead of the average and the standard deviation, to mitigate the effect of large fluctuations of the resolved flux that can be caused by the occasional occurrence, in some realizations, of a powerful source located near the Sun.}

In the case of point-like sources, the predictions of our model are compatible, within $2\sigma$, with the observed values in all the considered regions.
In particular, the agreement between the expected and measured flux from resolved sources at $50$ TeV is remarkable.
This result is not obvious since our synthetic population was built based on the H.G.P.S. catalog \cite{H.E.S.S.:2018zkf}. 
Our finding basically implies that the same synthetic population that accounts for H.E.S.S. observations, at an average source energy of few TeVs, can also account for the results obtained by LHAASO-KM2A at 50 TeV.
One important ingredient in our simulation is the average source spectrum, parameterized as a power-law with spectral index $2.4$ and $E_{\rm cut}=100$ TeV.
The adopted cutoff energy is sufficiently high to maintain agreement with H.G.P.S. observations and, at the same time, sufficiently low to avoid overestimating the number of bright (and hard) sources in the LHAASO-KM2A catalog.
All the above considerations also hold in the case where the source size is fixed at 40 pc (see Tab.~\ref{tabLHAASO40}).

These results can be considered as validation of our synthetic population, a necessary condition to ensure that estimating its contribution to the unresolved gamma-ray flux is meaningful.

\begin{table}
\centering
\scriptsize
\renewcommand{\arraystretch}{1.4} %
\begin{tabular}{l|ccccccc}
\hline
\hline
 &  & $N_{\rm R}$ & $\varphi_{\rm R}$ & $\varphi_{\rm UNR}$ & $\varphi_{\rm UNR,H}$ \\
 \hline
$15^{\circ}<l<235^{\circ}, |b|<5^{\circ}$ & MC & $84_{-9}^{+9}$ & $1.69_{-0.43}^{+0.62}\times 10^{-14}$ & $2.82_{-0.14}^{+0.15}\times 10^{-15}$ & $-$\\
 & KM2A & $65$ & $1.51\times 10^{-14}$ & $-$ & $-$ \\
 \hline 
 
$15^{\circ}<l<125^{\circ}, |b|<5^{\circ}$ & MC & $72_{-8}^{+8}$ & $1.32_{-0.33}^{+0.37}\times 10^{-14}$ & $2.56_{-0.16}^{+0.14}\times 10^{-15}$ & $2.23_{-0.36}^{+0.34}\times 10^{-16}$\\
 & KM2A & $55$ & $1.38 \times 10^{-14}$ & $-$ & $-$ \\
 \hline
$125^{\circ}<l<235^{\circ}, |b|<5^{\circ}$ & MC & $12_{-4}^{+4}$ &  $2.82_{-1.1}^{+1.8}\times 10^{-15}$ & $2.53_{-0.35}^{+0.46}\times 10^{-16}$ & $2.08_{-0.34}^{+0.49}\times 10^{-16}$ \\
 & KM2A & $10$ & $1.30 \times 10^{-15}$ & $-$ & $-$ \\
 \hline
\end{tabular}
\caption{\em The table compares, in the different spatial windows, defined in the leftmost column, the model predictions for our point-like source population and the LHAASO results. The rows with the indication MC (KM2A) report model (observation) results. The first two columns report the number, $N_{\rm R}$, of detectable (MC) vs detected (KM2A) sources and their total flux, $\varphi_{\rm R}$. The rightmost columns report the differential flux expected at $50$ TeV from unresolved sources, in the entire window, $\varphi_{\rm UNR}$, and limited to the masked region, $\varphi_{\rm UNR, masked}$. Fluxes are expressed in $\,TeV^{-1}\,cm^{-2}\,s^{-1}$.
The Monte Carlo (MC) predictions for the source number and fluxes are derived as the average and median values, respectively, over 100 random realizations of the PWN source population. The errors on the number represent the variance, while the lower and upper errors on the fluxes correspond to the 25 and 75 percentile, respectively.
The KM2A rows are obtained from the KM2A catalog \cite{LHAASO:2023cat}.}
\label{tabLHAASO}
\end{table}

\begin{table}
\centering
\scriptsize
\renewcommand{\arraystretch}{1.4} %
\begin{tabular}{l|ccccccc}
\hline
\hline
 &  & $N_{\rm R}$ & $\varphi_{\rm R}$ & $\varphi_{\rm UNR}$ & $\varphi_{\rm UNR,H}$ \\
 \hline
$15^{\circ}<l<235^{\circ}, |b|<5^{\circ}$ & MC & $69_{-7}^{+7}$ & $2.28_{-0.56}^{+0.74}\times 10^{-14}$ & $7.17_{-0.45}^{+0.37}\times 10^{-15}$ & $-$\\
 & KM2A & $65$ & $1.51\times 10^{-14}$ & $-$ & $-$ \\
 \hline
$15^{\circ}<l<125^{\circ}, |b|<5^{\circ}$ & MC & $62_{-7}^{+7}$ & $1.76_{-0.45}^{+0.47}\times 10^{-14}$ & $6.10_{-0.41}^{+0.37}\times 10^{-15}$ & $8.19_{-1.64}^{+1.74}\times 10^{-16}$\\
 & KM2A & $55$ & $1.38 \times 10^{-14}$ & $-$ & $-$ \\
 \hline
$125^{\circ}<l<235^{\circ}, |b|<5^{\circ}$ & MC & $7_{-3}^{+3}$ &  $4.21_{-1.8}^{+3.4}\times 10^{-15}$ & $1.05_{-0.27}^{+0.25}\times 10^{-15}$ & $8.32_{-1.99}^{+2.49}\times 10^{-16}$ \\
 & KM2A & $10$ & $1.30 \times 10^{-15}$ & $-$ & $-$ \\
 \hline
\end{tabular}
\caption{\em Same as Tab.~\ref{tabLHAASO} but for extended sources with size fixed at $40$ pc.}
\label{tabLHAASO40}
\end{table}

\section{Results}
\label{sec:Results}

\begin{figure*}
\begin{center}
\subfigure[]{\includegraphics[width=0.45\textwidth]{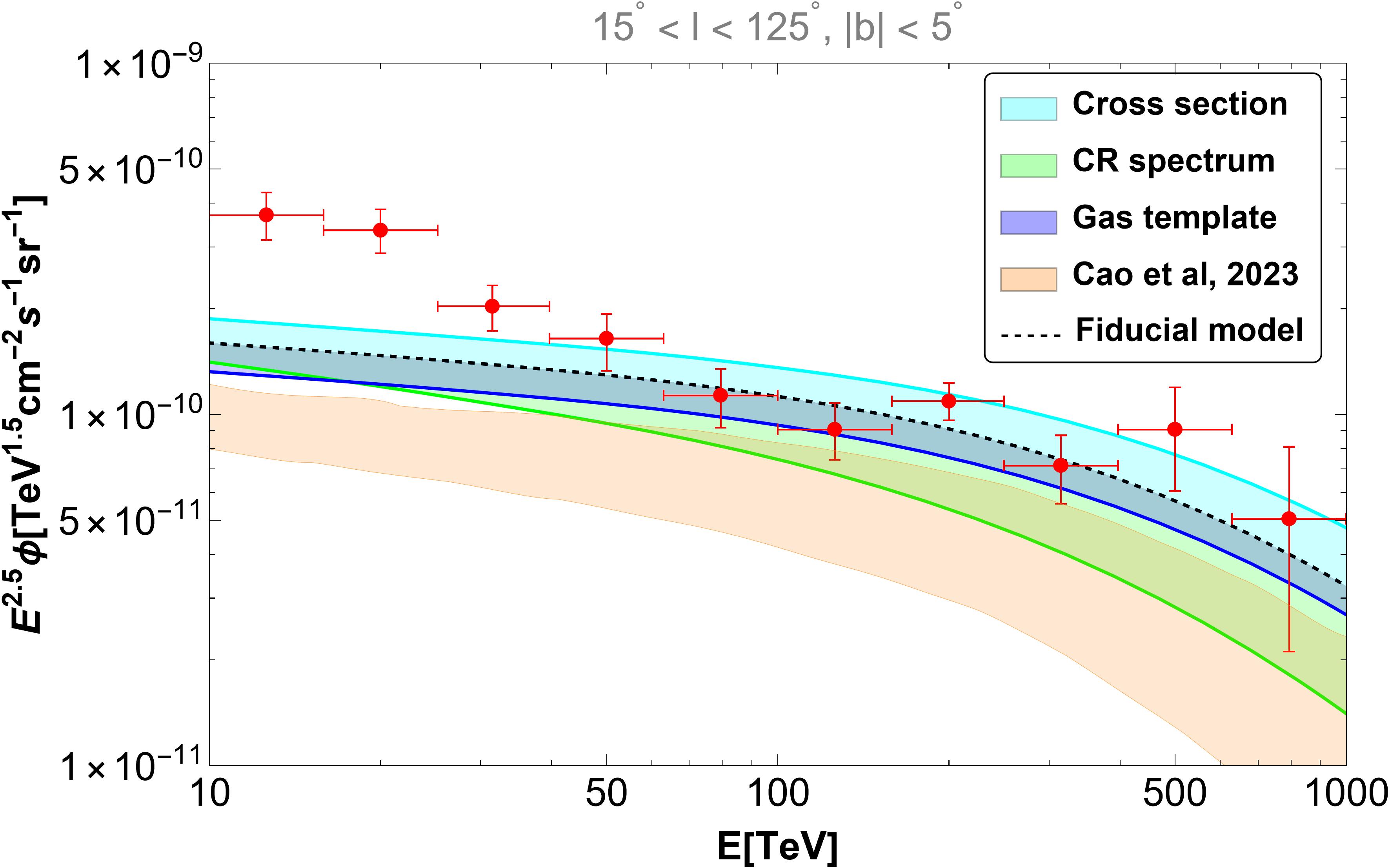}}
\subfigure[]{\includegraphics[width=0.45\textwidth]{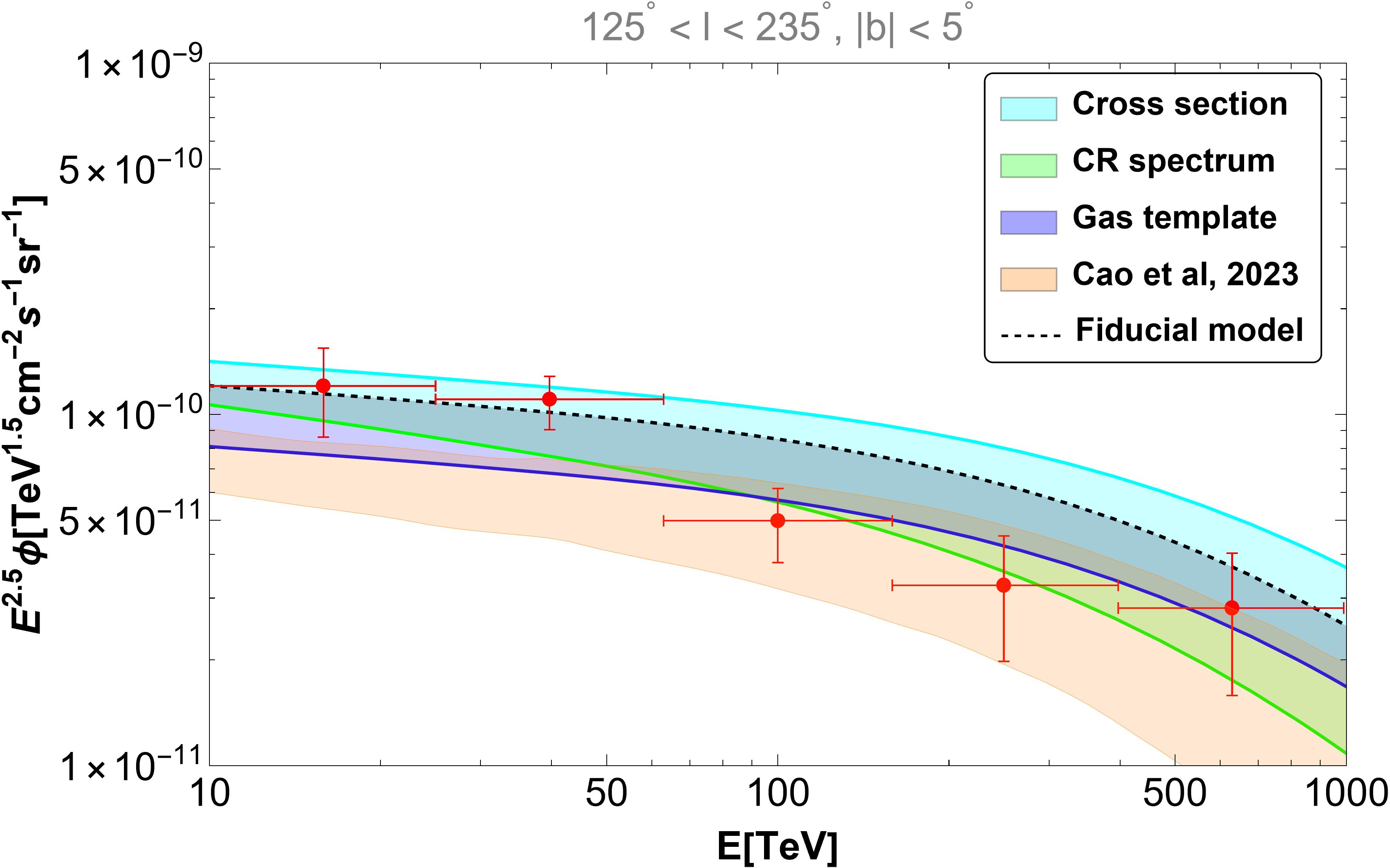}}
\caption{\small\em Differential energy spectra of diffuse $\gamma-$rays from the Galactic plane in the two angular regions probed by the LHAASO detector. Inner region in panel (a) and outer region in panel (b). Red data points are the measurements provided by LHAASO \cite{LHAASO:2023diff}. The dashed black line represents our fiducial model for the diffuse emission. The cyan, green, and blue shaded bands represent the variation with respect to the fiducial model related to a change of the cross-section, the CR spectrum, and the gas template assumptions, respectively.
The orange band represents the theoretical predictions used by the LHAASO collaboration \cite{LHAASO:2023diff}.
}
\label{fig:LHAASOstandard}
\end{center}
\end{figure*} 

\subsection{Diffuse emission}
\label{sec:results diffuse}
Fig.~\ref{fig:LHAASOstandard} shows our results for the Galactic gamma-ray diffuse emission in the inner (panel (a)) and outer (panel (b)) regions probed by LHAASO compared with the corresponding experimental results.
In each panel, the dashed black line shows the predictions of our fiducial model. 
For both regions, we have applied the mask adopted by LHAASO to eliminate the contribution produced by resolved sources.
The cyan, green, and blue shaded bands show the deviations from the fiducial model produced by variations of the assumed $pp$ cross-section, CR spectrum, and gas template, respectively.
%
In particular, we note an uncertainty of up to $\sim 20\%$ associated with the cross section: the Pythia 8.18 parametrization (upper bound of cyan band), predicts more gamma-rays than the fiducial model, which adopts the AAFRAG parametrization.
%
By looking at the blue shaded area, it can be noticed that the dust provides a smaller column density than the gas obtained from the GALPROP maps: this results in a gamma ray flux that is $\sim 20\%$ ($\sim 50\%$) smaller in the LHAASO inner (outer) region. 
%
The discrepancy due to the different CR proton spectra as measured by  IceTop \cite{IceCube:2019hmk} and KASCADE \cite{Apel:2013uni} is rather small at $10$ TeV but increases at larger energies, reaching up to a factor of $\sim 2$ at PeV energies. 

For the sake of readability, in Fig.~\ref{fig:LHAASOstandard}, we don't display the uncertainties related to the CR spatial distribution.
After masking, the difference between the two spatial assumptions considered in this work, i.e., $g({\bf r})= g_{\rm snr}({\bf r})$ and $g({\bf r})=1$, is negligible in the inner region (less than $10\%$), while in the outer region the former results into $\sim 20\%$ less signal than the latter, as can be expected due to a decreasing number of SNRs in the Galaxy outer region. 

Panel (b) in Fig.~\ref{fig:LHAASOstandard} clearly shows that the prediction of our fiducial model is compatible with LHAASO results in the outer region and that the relatively small discrepancies 
can be easily accommodated within the uncertainty ranges discussed above.
In the inner region (panel (a) of the same Figure), our calculations are still compatible with experimental data above $\sim$ 50 TeV, without the need to assume CR spectral hardening in the inner Galaxy and/or a non-negligible contribution from unresolved sources. 
However, the first three data points, below 50 TeV, are above expectations, even after accounting for all the uncertainties, suggesting that a possible additional contribution from unresolved sources could indeed be relevant in this energy band.
The above conclusions are then different from those put forward by the LHAASO collaboration, according to which the measured fluxes exceed the theoretical expectation by a factor $\sim 2-3$. 
We note, however, that the theoretical model used in that work accounted only for uncertainties in the CR spectrum, while their assumptions for the cross section (AAFRAG) and gas distribution (derived from the dust) both tend to minimize the diffuse flux.
%
%
For reference, we reported the theoretical predictions used by the LHAASO collaboration as an orange band in Fig.~\ref{fig:LHAASOstandard} (see Fig.~2 of \cite{LHAASO:2023diff}).

\subsection{Hardening of the CR spectrum}
\label{Sec:hardening}
Even if our results considerably weaken the evidence for an excess of the observed signal with respect to standard predictions, it is useful to test the effects of a progressive hardening of the CR spectrum toward the Galactic center in order to understand whether such a scenario is potentially relevant and/or can be constrained by LHAASO data.
We first remark that the outer region probed by LHAASO is not affected by this possibility, simply because the longitude range $125^\circ < l < 250^\circ$ is far apart from the region for which the hardening was originally invoked based on {\it Fermi}-LAT data.
%
As a consequence, we only evaluate the impact of the hardening in the inner region, where it can indeed produce a sizable enhancement of the emission.
%
However, also in this region, the masking procedure adopted by LHAASO considerably lessens the effect.
This is because the sky region defined by $l\le 80^\circ$ and $|b|< 1^\circ$, 
is almost completely vetoed by the LHAASO mask, filtering out the contribution of most internal regions of the Galaxy where the effect of CR hardening is more relevant.

\begin{figure}
\begin{center}
\includegraphics[width=0.65\textwidth]{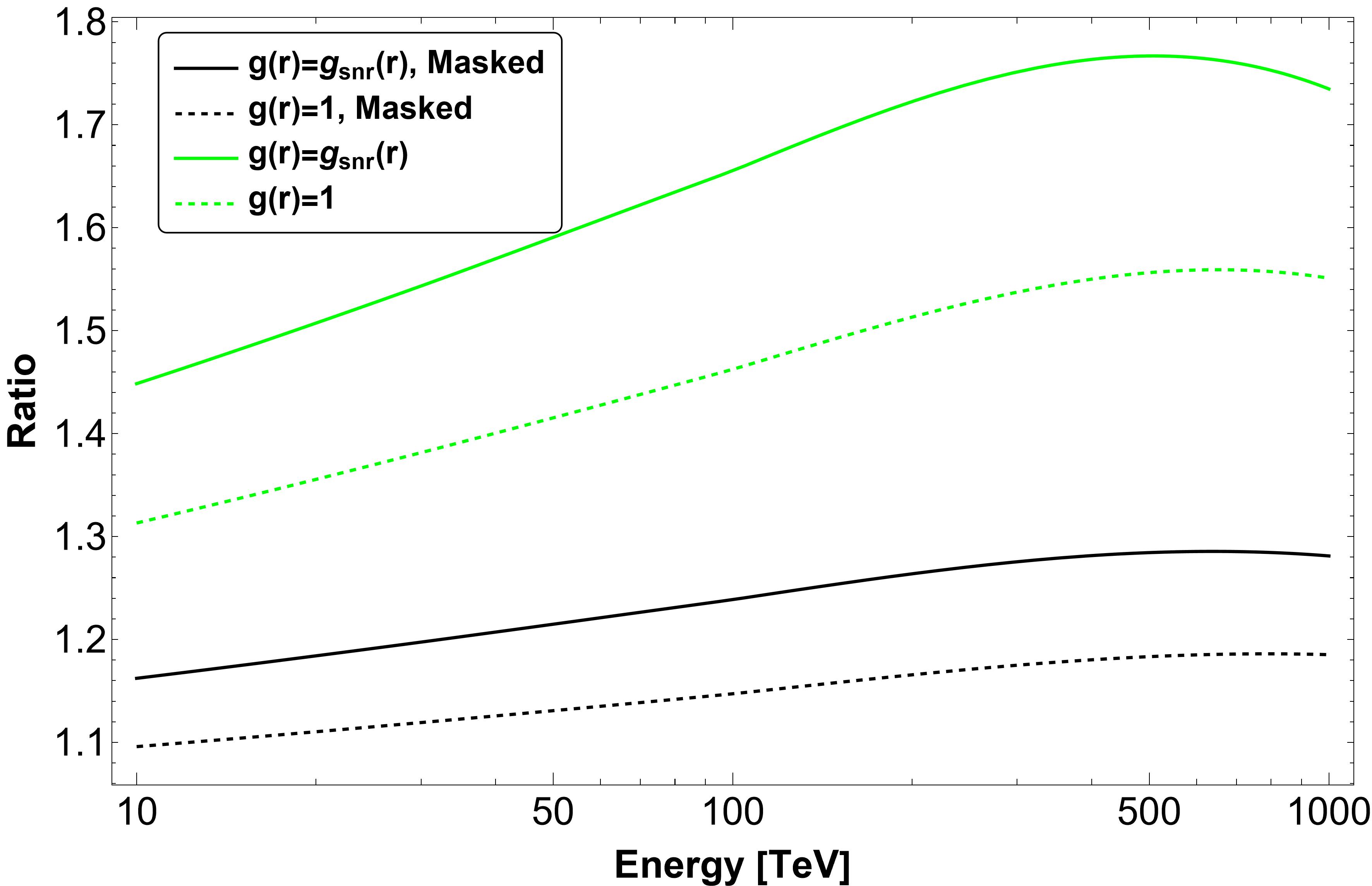}
\caption{\small\em Ratio, in the inner LHAASO region,  between the total diffuse emission computed including the galactic center hardening and that computed in our fiducial case.
Green and black lines refer to the flux before and after applying the LHAASO masks, while the dashed and solid lines identify the cases with spatially uniform and non-uniform CRs respectively.
}
\label{fig:Masking effect}
\end{center}
\end{figure}

In Fig.~\ref{fig:Masking effect}, we show the enhancement of the diffuse gamma-ray emission in the inner region resulting from the CR spectral hardening, compared to our fiducial case.
The black and green lines are obtained with and without applying the LHAASO mask, respectively.
The enhancement factor is independent of the adopted cross-section, gas template, and CR spectrum.
It depends, however, on the assumed CR spatial distribution, as it can be appreciated by comparing results obtained with the two assumptions $g({\bf r}) = g_{\rm snr}({\bf r})$  and $g({\bf r}) = 1$ considered in this paper.
Clearly, one has larger effects in the first case because this corresponds to concentrating more CRs in the central region of the Galaxy (where CR sources are more abundant) rather than at its edges.
In particular, for $g_{\rm snr}({\bf r})$, the CR hardening hypothesis predicts $44\%$ ($76\%$) larger gamma-ray emission than the standard case at $10$ TeV ($500$ TeV), see green solid line in Fig.~\ref{fig:Masking effect}.
However, even in this most favorable case, the enhancement is significantly suppressed once the mask is applied, decreasing to $16\%$ ($28\%$) at $10$ TeV ($500$ TeV) (see black solid line).
In conclusion, the effect of CR hardening is reduced by the masking procedure to a level comparable to or smaller than the other uncertainties discussed in the previous section, making it challenging to drive any firm conclusions based solely on the comparison of theoretical predictions with the LHAASO data.

\begin{figure*}
\begin{center}
\subfigure[]{\includegraphics[width=0.45\textwidth]{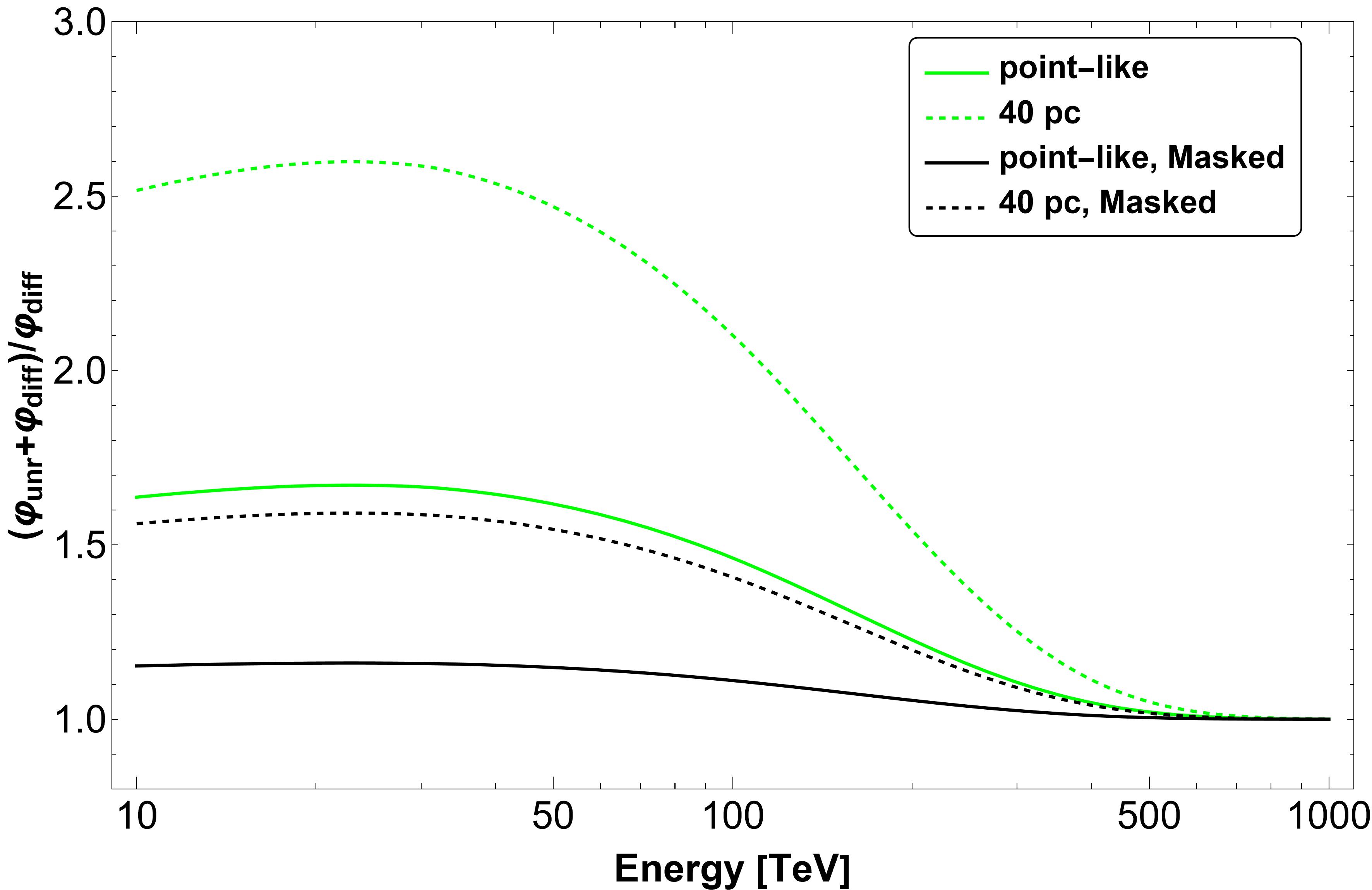}}
\subfigure[]{\includegraphics[width=0.45\textwidth]{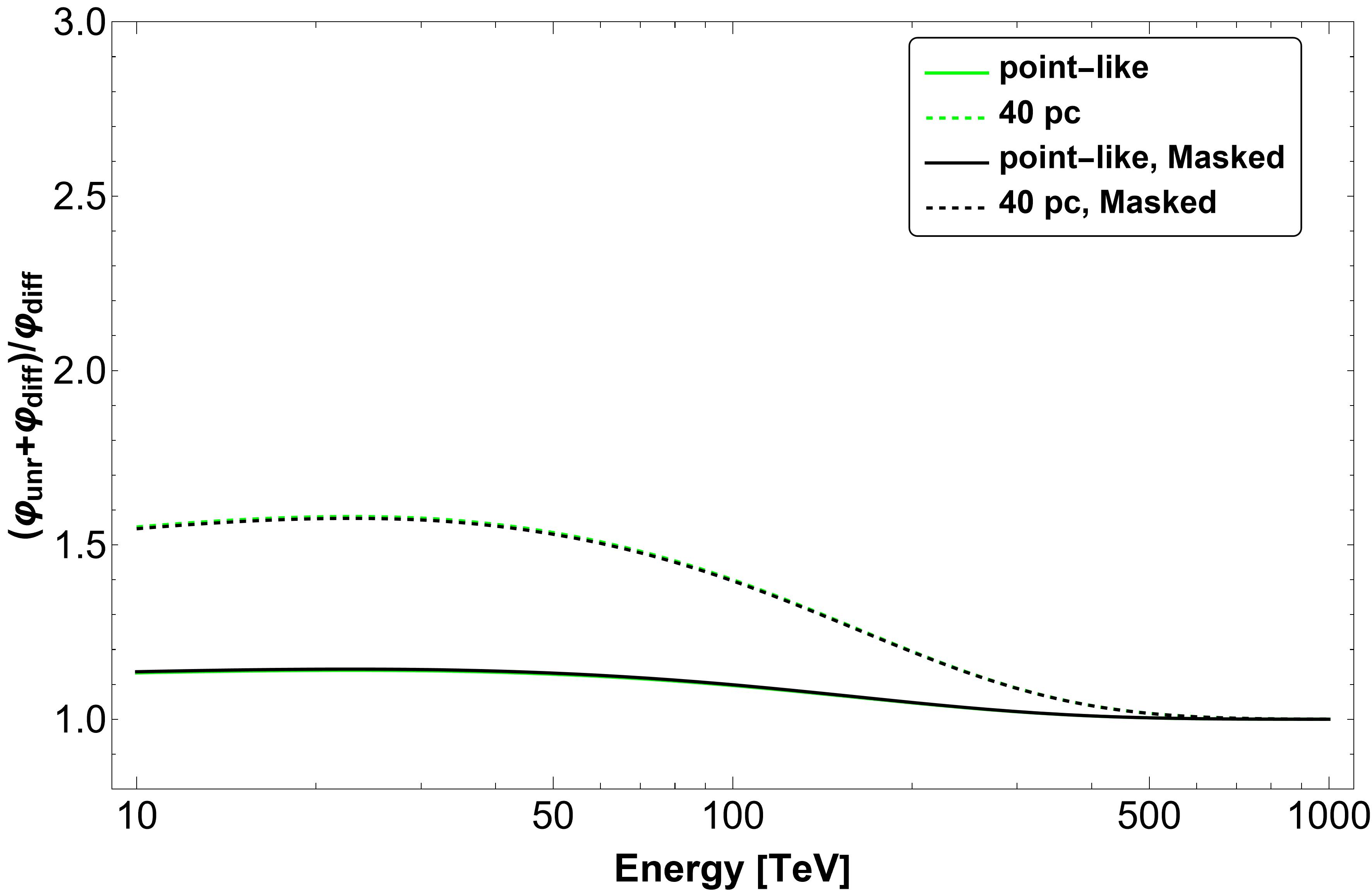}}
\caption{\small\em 
Ratio of the total diffuse signal (i.e., truly diffuse + unresolved sources) to the truly diffuse emission alone as a function of energy. The predictions for the inner region are shown in panel (a) and for the outer region in panel (b). 
Green lines are for total fluxes while black lines show the effect of the mask. Solid and dashed lines are obtained for point-like sources and for the source size fixed to $40$ pc, respectively. 
}
\label{fig:Masking effect source}
\end{center}
\end{figure*}

\begin{figure*}
\begin{center}
\subfigure[]{\includegraphics[width=0.45\textwidth]{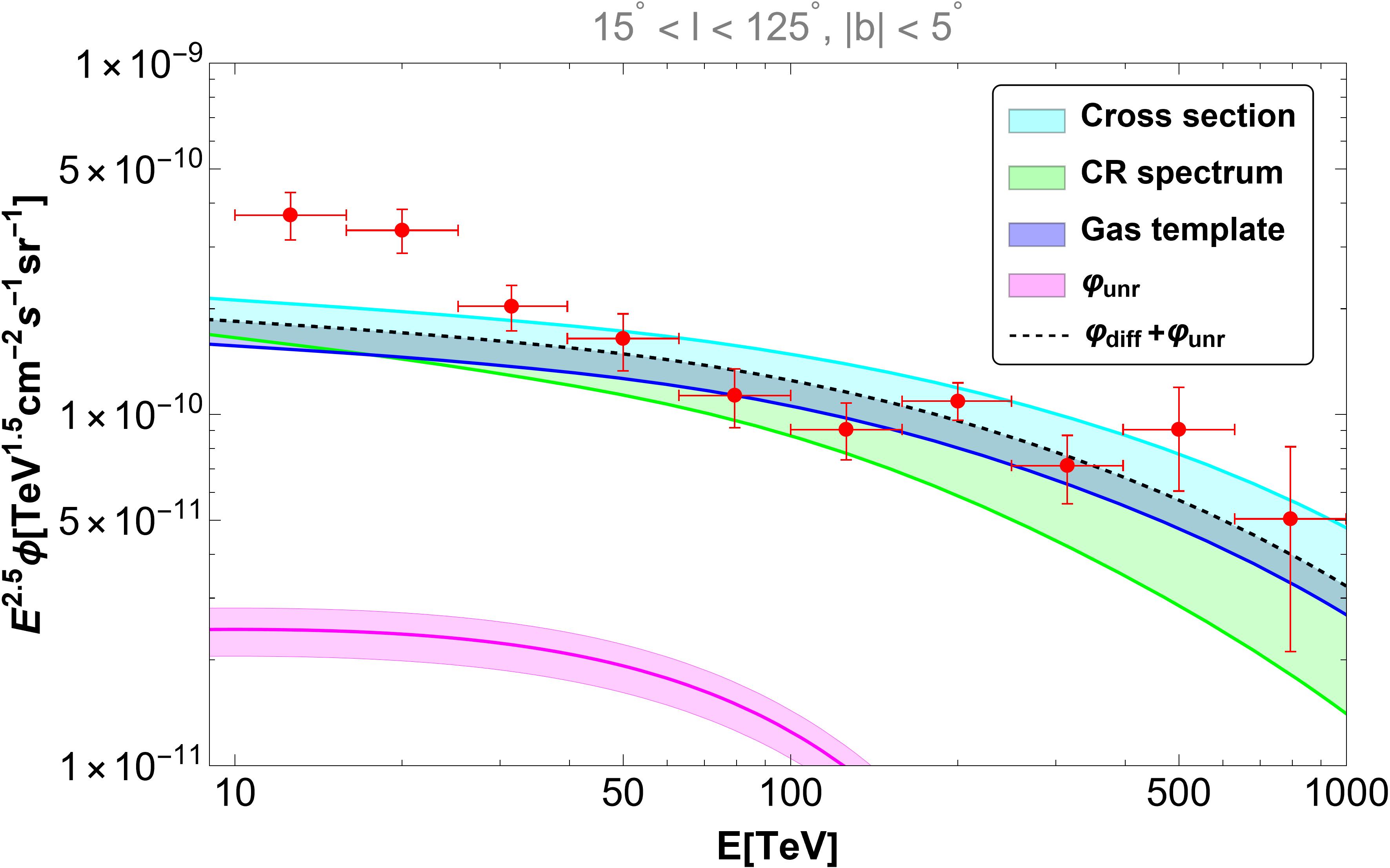}}
\subfigure[]{\includegraphics[width=0.45\textwidth]{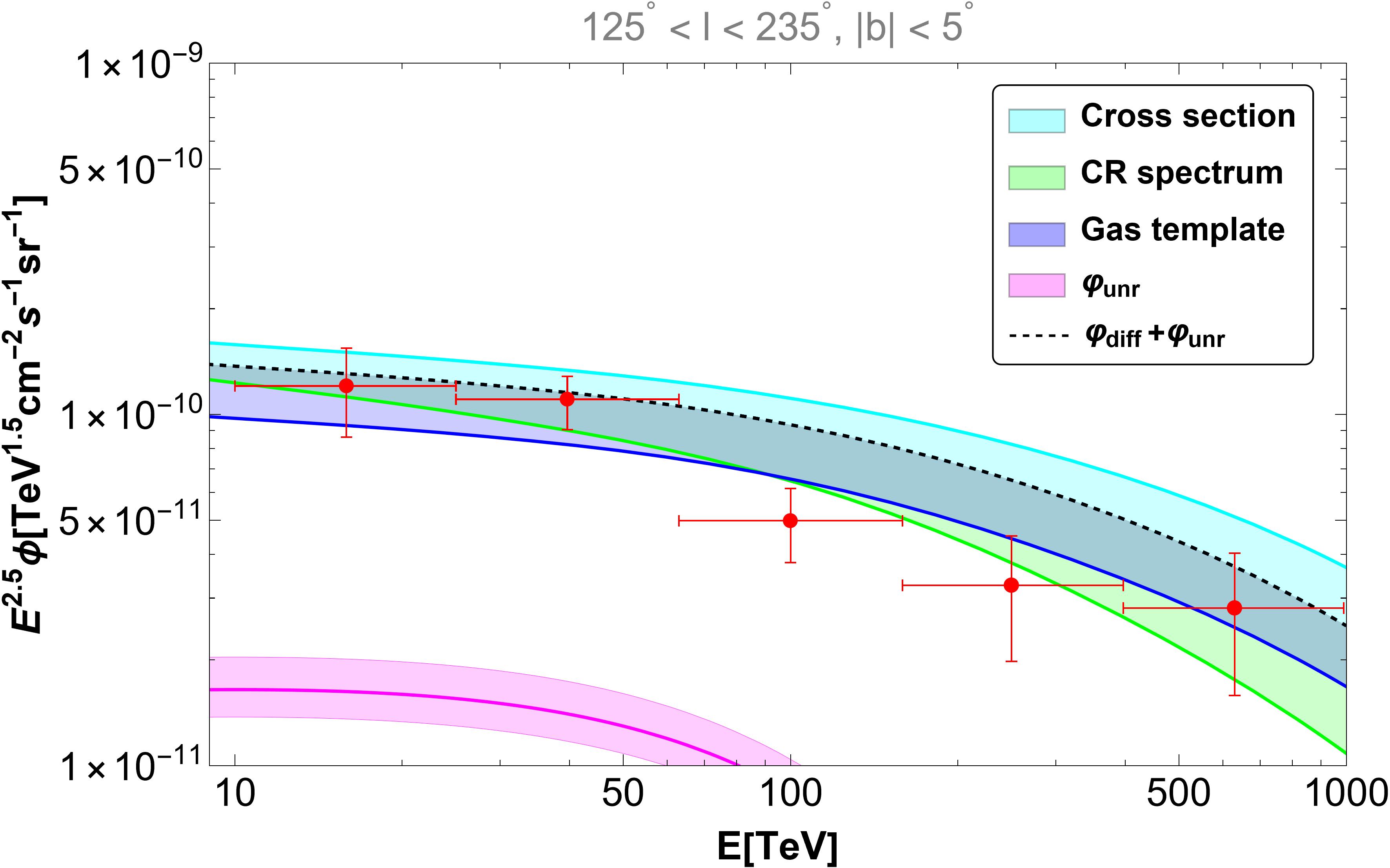}}
\caption{\small\em 
Same as Fig.~\ref{fig:LHAASOstandard} but with the dashed black line representing the total diffuse emission obtained as the sum of our fiducial model plus the median value for the unresolved sources. 
The cyan, green, and blue shaded bands represent the fiducial model's variation with respect to different assumptions on the cross-section, the CR spectrum, and the ISM, respectively.
The solid thick magenta line represents the median value of the contribution from point-like unresolved sources while the lower (upper) bound of the magenta band represents the first (third) quartile.
}
\label{fig:LHAASOstandardPL}
\end{center}
\end{figure*}

\begin{figure*}
\begin{center}
\subfigure[]{\includegraphics[width=0.45\textwidth]{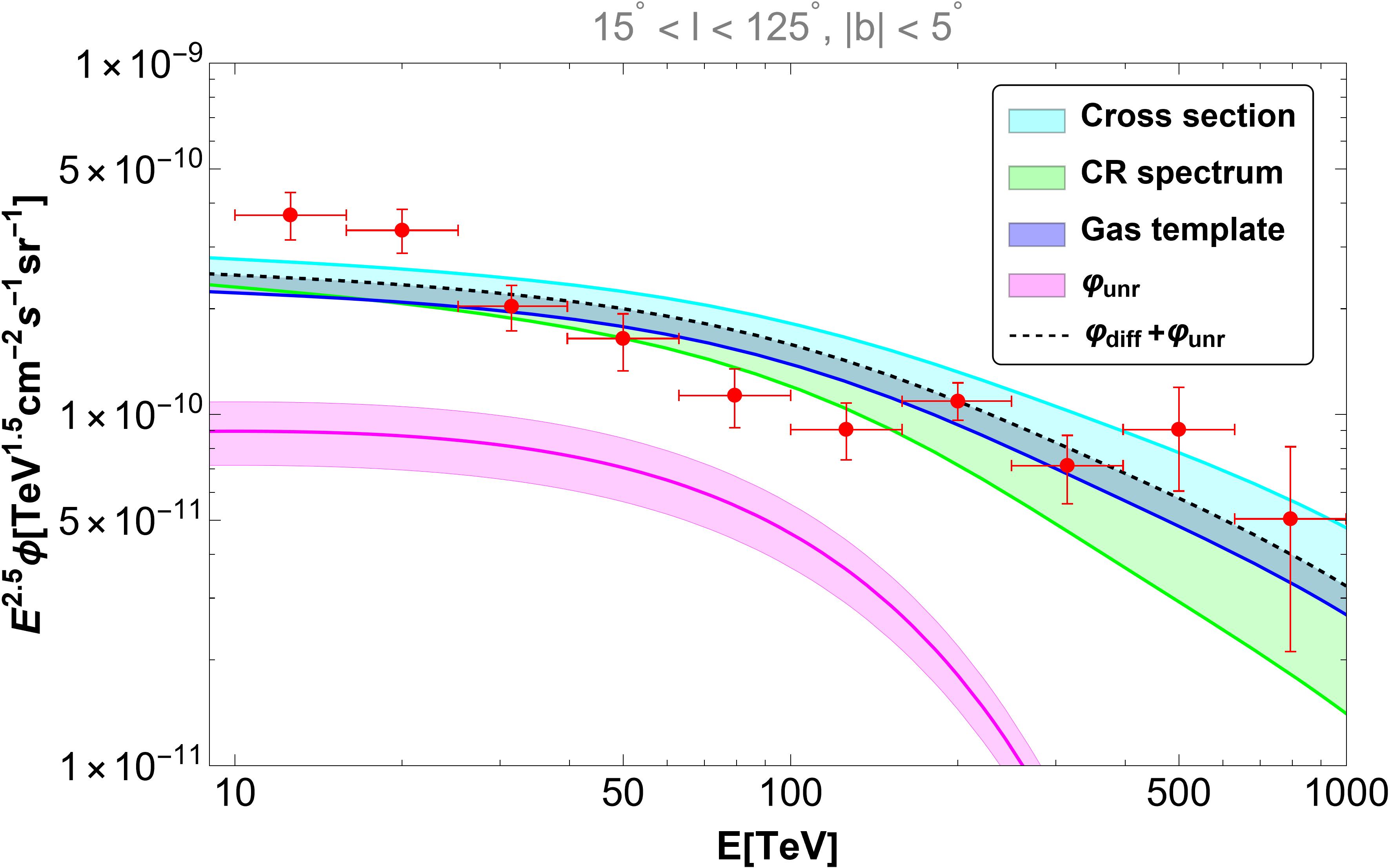}}
\subfigure[]{\includegraphics[width=0.45\textwidth]{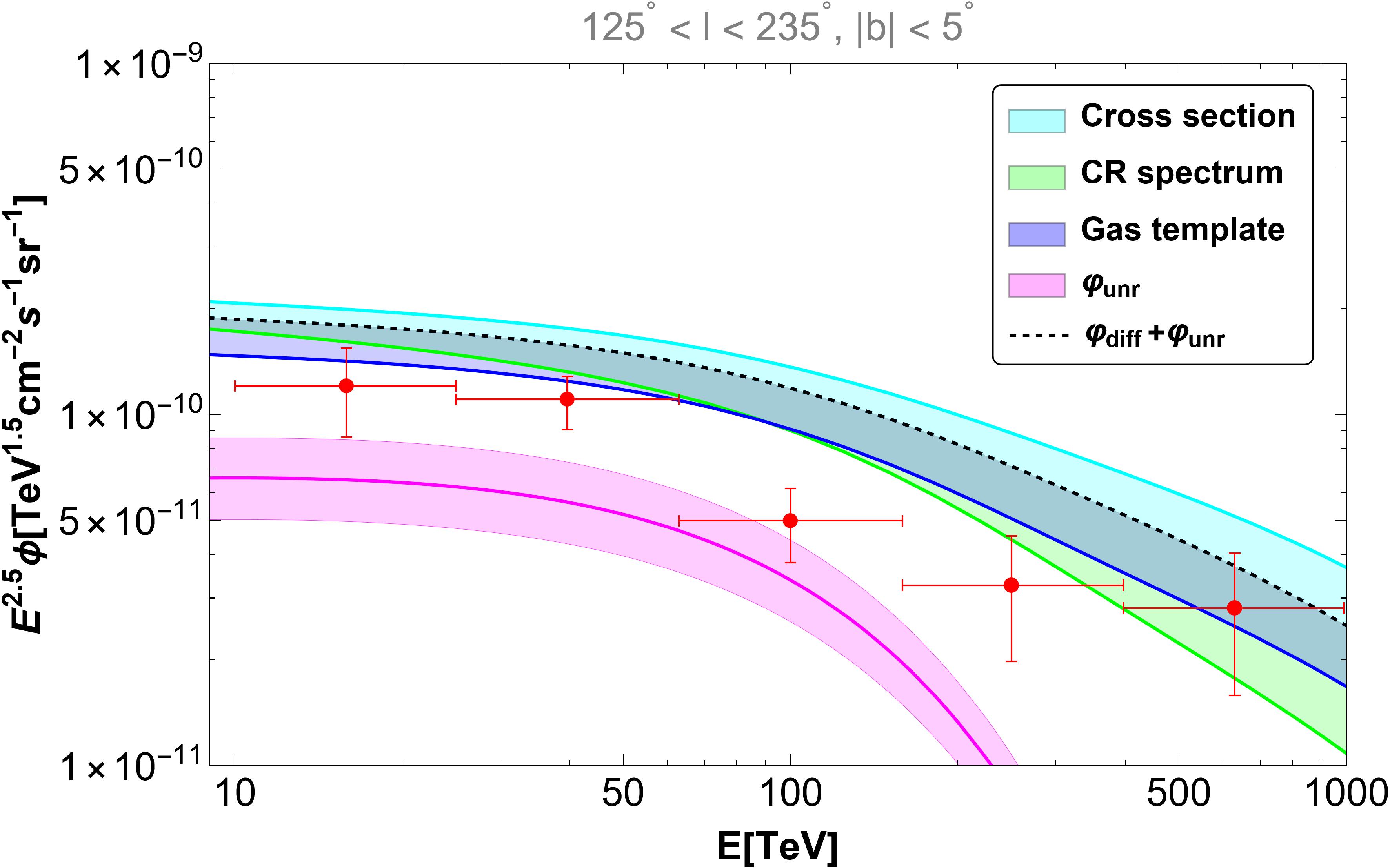}}
\caption{\small\em Same as Fig.~\ref{fig:LHAASOstandardPL} but for source size fixed to 40 pc.
}
\label{fig:LHAASOstandard40pc}
\end{center}
\end{figure*}

\subsection{Unresolved source contribution}
\label{sec:unresolved}

We finally use our synthetic population to estimate the contribution of unresolved sources in the LHAASO energy range and to assess their relevance for the interpretation of experimental results.
By comparing columns 3 and 4 in Tab.~\ref{tabLHAASO} and Tab.~\ref{tabLHAASO40}, it becomes clear that the masking procedure significantly suppresses the unresolved source flux, especially in the inner region.
In particular, at $50$ TeV, the mask suppresses $\sim 90\%$ of the unresolved source contribution in the inner region and $\sim 20\%$ in the outer region.
The reduction factors are approximately equal for point-like and extended sources.

To better assess the relevance of unresolved sources, we show in Fig.~\ref{fig:Masking effect source} the ratio between the total diffuse signal, obtained as the sum of truly diffuse emission and unresolved sources contribution, and the truly diffuse emission alone as a function of energy. 
The green and black lines are obtained before and after applying the LHAASO mask, respectively.
One can clearly see from Fig.~\ref{fig:Masking effect source} that the unresolved source contribution is potentially very relevant before applying the mask in the LHAASO inner region. 
In particular, it can produce an enhancement of the gamma-ray flux at $50$ TeV by a factor $\sim1.6$ ($\sim2.5$) for point-like (extended) sources (see dashed and solid green lines in panel (a) of Fig.~\ref{fig:Masking effect source}).
However, after the mask is applied, the enhancement factor is significantly reduced: the mask not only suppresses the source contribution but also the truly diffuse emission that is reduced by about $\sim 60\%$ in the inner region and $\sim 20\%$ in the outer region.

The final result is that, in the case of point-like sources (solid lines), the unresolved source contribution significantly drops in both regions where the diffuse emission is measured.
More precisely, in both regions, it corresponds to about $15\%$ of the fiducial model predictions at $50$ TeV (see black solid lines in Fig.~\ref{fig:Masking effect source}).
This is illustrated more clearly in Fig.~\ref{fig:LHAASOstandardPL}, where a dashed black line shows the expected total diffuse emission compared with LHAASO measurements.
This is obtained as the sum of the median value of the flux from point-like unresolved sources, after masking (magenta thick line) and our fiducial model.
On the other hand, as expected, the unresolved source contribution is larger in the case of extended sources because of the degraded detector sensitivity.
In particular, the flux at 50~TeV due to unresolved extended sources, after the masking procedure, corresponds to $\sim 55\%$ of the fiducial model prediction (see dashed black line in Fig.~\ref{fig:Masking effect source}) and provides up to the $\sim 35\%$ of the total diffuse signal.
This case, however, has to be considered as an upper limit that could potentially provide a too-large total signal in the outer LHAASO observation window where the truly diffuse emission alone already saturates the observed signal. 
This becomes clear when looking at panel (b) of Fig.~\ref{fig:LHAASOstandard40pc}, where the total flux, obtained as the sum of diffuse emission and extended unresolved source contribution, overshoots the LHAASO data at $100$ TeV.

We finally note that the enhancement factors obtained in the outer region, see panel (b) of Fig.~\ref{fig:Masking effect source}, before (green lines) and after applying the mask (black lines) are approximately the same.
This is due to the fact that the mask acts in the same way on diffuse emission and unresolved sources, by reducing both components by a similar factor (equal to about $\sim 20\%$) that cancels out when computing ratios.

\section{Conclusions}
\label{sec:Conclusions}
In this work, we compare the ultra-high-energy data provided by LHAASO with an improved modeling of the total diffuse emission. 
We calculate the diffuse emission produced by the interaction of CRs with the ISM under different assumptions on the CR spectrum, based on the results of different CR experiments, on the $pp$ inelastic cross-section, and on the gas tracers, and argue that these systematic uncertainties are not negligible. 
We show that the LHAASO data are reasonably well explained by the hadronic diffuse emission in the outer region without the need for a dominant contribution from unresolved sources. This statement remains valid in the inner region above $\sim$ 50 TeV.

We test the effect of a possible spatial dependence of the CR spectral index, i.e., a hardening, \cite{Acero:2016qlg, Yang:2016jda, Pothast:2018bvh} on the predicted gamma-ray diffuse emission.
We find that, in the inner region, after applying the mask adopted by the LHAASO collaboration, the hardening effect produces a variation of at most $30\%$ compared with the fiducial model. 
Since the sources are primarily distributed along the Galactic Plane and the LHAASO masks are chosen to be very large, $5$ times larger than the size of the sources themselves, the exclusion regions mostly cancel the inner part of the Galaxy, where the hardening effect is expected to be most pronounced.
As a consequence, it is not possible to disentangle between standard and inhomogeneous diffusion in our Galaxy based on the released LHAASO data alone.

Finally, we discuss and quantify the expected contribution due to unresolved TeV sources in the LHAASO data.
We calculate the contribution of unresolved sources to the LHAASO measurement using a population study \cite{Cataldo:2020qla} of TeV sources included in the H.G.P.S. \cite{H.E.S.S.:2018zkf}. In our analysis we consider two scenarios: one where the sources are point-like, as expected for PWNe, and an extreme case where they have a fixed size of 40 pc, to assess the impact of source size on the predicted unresolved flux. 
In the case of point-like sources, we find that after applying the LHAASO mask, their contribution is marginal. 
This is not surprising because the largest contribution to the flux of unresolved sources is expected from regions of the sky that are masked in this specific analysis.
The unresolved flux that survives the cut is almost negligible and provides a contribution of about $15\%$ to the hadronic diffuse emission at $50$ TeV in both sky regions. This result is consistent with the findings of the recent study by \cite{Kaci:2024lwx}.
This confirms our findings about the LHAASO data being compatible with the truly diffuse emission in the full energy range in the outer region and above $\sim$ 50 TeV in the inner region.
On the other hand, in the case of extended sources with a physical size of 40 pc, the contribution of unresolved sources with respect to the truly diffuse emission is large even after the masking procedure, accounting for about $55\%$ of it at $50$ TeV. However, this extreme scenario leads to an overshoot of the LHAASO data in the outer region. 
The scenario we considered is indeed extreme, both in terms of source size and in terms of source abundance (we assumed every PWN in our synthetic population to be surrounded by a 40 pc halo), but our result probes the possibility of using ultra-high-energy data to constrain the existence and size of regions of suppressed diffusion around sources. 
Indeed, in the inner region, the contribution of unresolved extended sources could help explain the LHAASO data below {\bf $\sim$ 50 TeV}, which remains \sout{slightly} above our predictions. 
Finally, it is worth mentioning that while this analysis was focused on the possible contribution of unresolved PWNe and pulsar haloes, another source class that is important to consider in this respect is that of stellar clusters. Due to the large typical sizes of their wind-blown bubbles, where CRs are efficiently confined, these sources are extended, rather than point-like, and can provide an important contribution exactly in the region where it seems to be needed \cite{Menchiari:2024uce}.
To conclude, it is relevant to mention that diffuse emission data are also available at GeV energy from the Fermi-LAT analysis \cite{Zhang:2023ajh} (see Appendix.~\ref{App:Fermi})
However, a fair comparison with GeV data requires a proper parametrization of additional diffuse components not considered in this work, such as the bremsstrahlung and inverse Compton emission, which provide a non-negligible contribution at $few$ GeV.
Moreover, the contribution from unresolved sources needs to be evaluated in this energy range as well.
The unresolved sources in this work are tuned to high-energy data and specifically calculated for the LHAASO KM2A detector. In order to enable comparison, an appropriate parametrization of the spectra of unresolved sources at lower energies is required.
In the future, an analysis of the GeV energy range could provide valuable additional insights to this work.

\section{Acknowledgements}
\label{sec:acknowledgement}
The Authors are grateful to Andrew Taylor and Felix Aharonian for helpful discussions on the $pp$ cross-sections.
The work of VV is supported by the European Research Council (ERC) under the ERC-2020-COG ERC Consolidator Grant (Grant agreement No.101002352). 
G.P. research work is supported by the INAF Astrophysical fellowship initiative.
The work of G.P. and F.L.V. is partially supported by  grant number 2022E2J4RK "PANTHEON: Perspectives in Astroparticle and
Neutrino THEory with Old and New messengers" under the program PRIN 2022 funded by the Italian Ministero dell’Universit\`a e della Ricerca (MUR) and by the European Union – Next Generation EU. G.M. and E.A. are partially supported by INAF through Theory Grant 681 2022 ``{\it Star Clusters As Cosmic Ray Factories}'' and PRIN INAF 2019 ``From massive stars to supernovae and supernova remnants: driving mass, energy and cosmic rays in our Galaxy''. This work was partially supported by the European Union – NextGenerationEU RRF M4C2 1.1 under grant PRIN-MUR 2022TJW4EJ.
The work of S. M. acknowledges financial support from the Severo Ochoa grant CEX2021-001131-S funded by MCIN/AEI/ 10.13039/501100011033.

\paragraph{Note added.}  After the conclusion of this work, the LHAASO collaboration provided a new diffuse emission measurement based on the combined analysis of WCDA and KM2A detectors \citep{LHAASO:2024lnz}. The mask adopted in this new analysis differs from the previous one.
We provide a comparison of the new data with our model for the diffuse emission in Appendix \ref{App: WCDA}.
Our conclusion remains unchanged.

\bibliography{bibliography.bib}
\bibliographystyle{JHEP}

\appendix

\section{Cross-section}
\label{App:cross-sec}

\begin{figure*}[h!]
\begin{center}
\includegraphics[width=0.65\textwidth]{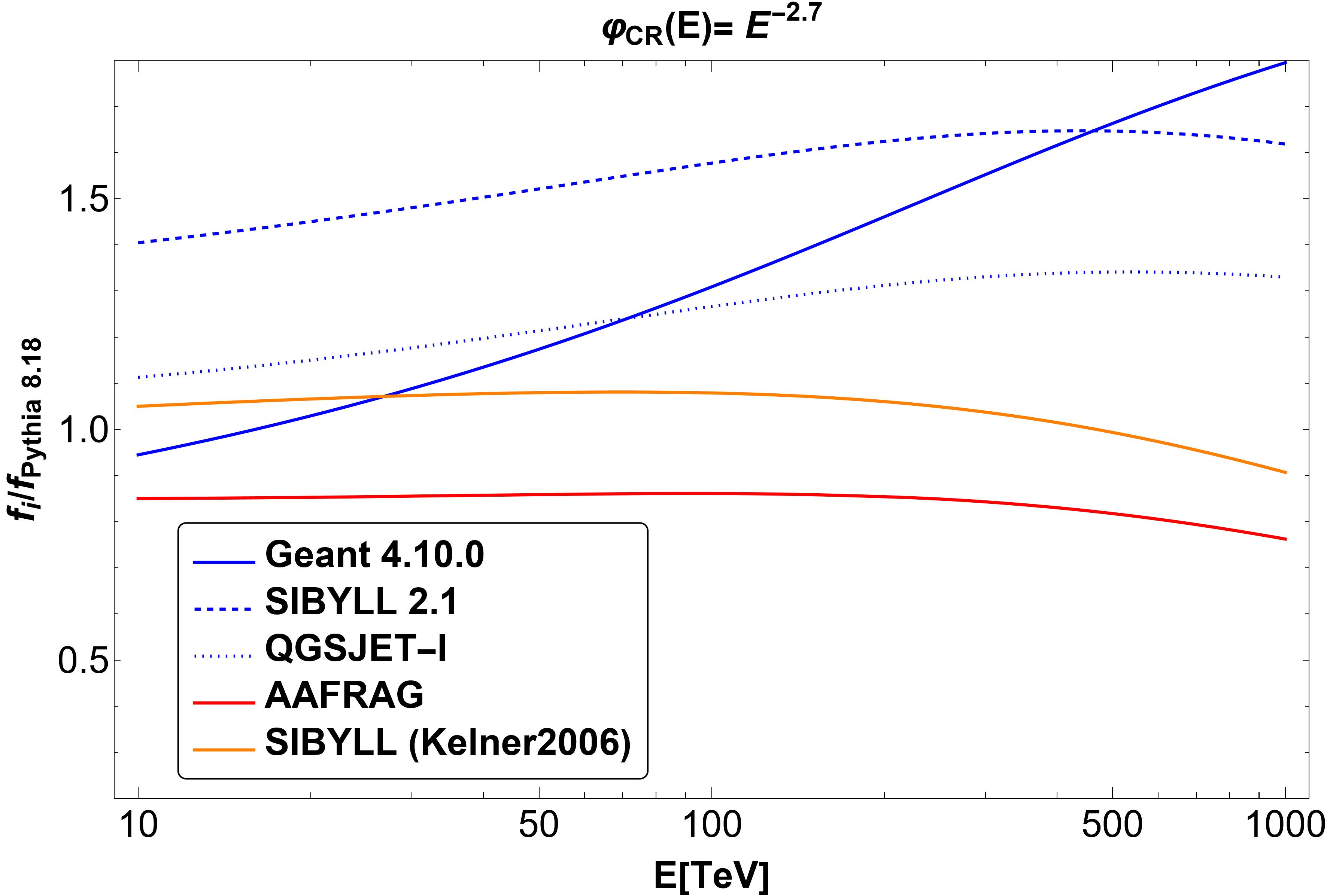}
\caption{\small\em  The ratio of gamma-ray emissivities for various parameterizations to that of the Pythia 8.18 parameterization \cite{Kafexhiu:2014cua} as a function of the gamma-ray energy. The solid, dashed, and dotted blue lines represent the Geant 4.10.0, SIBYLL 2.1, and QGSJET-I parameterizations from \cite{Kafexhiu:2014cua}, respectively. The red line represents AAFRAG and is taken from \cite{Kachelriess:2022khq}, while the orange line is the SIBYLL parameterization obtained in \cite{Kelner:2006tc}.
}
\label{fig:CrossSection}
\end{center}
\end{figure*}

In order to understand which of the considered parameterizations for the $pp$ cross-section provides the maximum and minimum gamma-ray signal, we calculate the gamma-ray emissivity defined as:

\begin{equation}
f_i(E_{\gamma})=\int_{E_\gamma}^{\infty} dE \frac{d\sigma_i(E,E_{\gamma})}{dE_{\gamma}}\varphi_{CR}(E)
\end{equation}
where $\frac{d\sigma_i(E,E_{\gamma})}{dE_{\gamma}}$ is the differential inelastic cross-section for the given parametrization $i$, $\varphi_{CR}(E)$ is the differential CR spectrum and the integral is performed over the CR energy $E$.
In Fig~\ref{fig:CrossSection} we show the ratio of gamma-ray emissivities for various parameterizations to that of the Pythia 8.18 parameterization \cite{Kafexhiu:2014cua}, as a function of gamma-ray energy. These ratios are calculated assuming $\varphi_{CR}(E) = E^{-2.7}$.
It can be seen that the maximum gamma-ray signal is produced by the SIBYLL 2.1 cross-section as parameterized by \cite{Kafexhiu:2014cua} (dashed blue line), while the minimum is given by AAFRAG \cite{Kachelriess:2022khq} (red solid line). The former produces a gamma-ray emissivity about $2$ times larger than the latter at $500$ TeV.
In general, the exact factor depends on the considered shape of the CR spectrum. However, the relative importance of the various parameterizations with respect to each other remains unchanged.
In the above consideration, we do not account for Geant 4.10.0, which is known to be not ideal for describing interactions at the highest energies.
Note that the AAFRAG parameterization is based on the QJSJET-II-04 model, which was specifically tuned to reproduce recent LHC data \citep{Kachelriess:2015wpa, Koldobskiy:2021nld}, while other parameterizations, such as SIBYLL 2.1, do not incorporate these updates \citep{Koldobskiy:2021nld}.
%
In \cite{Koldobskiy:2021nld}, significant differences were found in both the normalization and shape of the energy spectra when using the AAFRAG parameterization compared to other Monte Carlo generators, particularly in the region of large energy transfer. 
These discrepancies may arise from the fact that some parameterizations are based on pre-LHC event generators, whereas LHCf results on forward photon production support the adoption of the AAFRAG parameterization \cite{Koldobskiy:2021nld}.
For the above reason, we disregard the outdated SIBYLL 2.1 model from \cite{Kelner:2006tc, Kafexhiu:2014cua} and QGSJET-I from \cite{Kafexhiu:2014cua}. Instead, we adopt AAFRAG as our baseline model and consider Pythia 8.18 to account for cross-section uncertainties.


\section{Fit of the proton spectrum}
\label{App:Fit}
\begin{figure*}
\begin{center}
\includegraphics[width=0.65\textwidth]{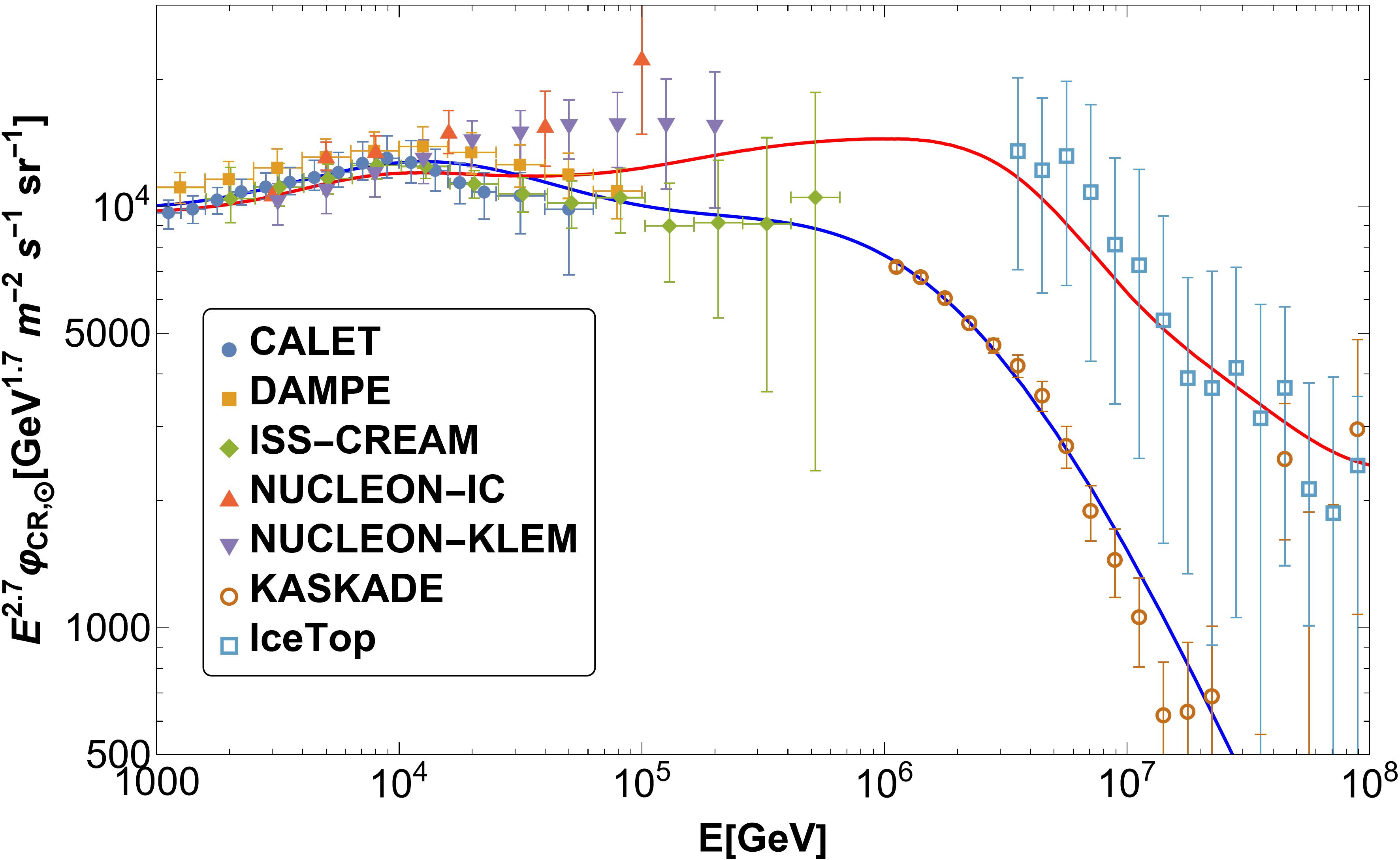}
\caption{\small\em The proton spectrum (blue line) obtained by fitting the data from CALET, DAMPE, ISS-CREAM, NUCLEON-IC, NUCLEON-KLEM and KASKADE. The proton spectrum obtained by \cite{Dembinski:2017} is shown with a red line. The error bars include both statistical and systematic uncertainties except in the case of KASKADE data, where only the statistical error is displayed.
}
\label{fig:KASKADEfit}
\end{center}
\end{figure*}

We perform a fit of the proton spectrum considering at PeV only the data from KASKADE \cite{Apel:2013uni}, while at lower energy, we combine the data from CALET \cite{CALET:2022vro}, DAMPE \cite{DAMPE:2019gys}, ISS-CREAM \cite{2022ApJISS-CREAM}, NUCLEON-IC, and NUCLEON-KLEM \cite{2019AdSpR_NUCLEON}.
The fit is performed using a power-law with two breaks:
\begin{eqnarray}
\nonumber
\phi_{\rm p} &=& K \left(\frac{E}{E_{0}}\right)^{-\alpha}\left[1+\left(\frac{E}{E_{b}}\right)^{1/\omega}\right]^{-(\alpha_1-\alpha)\omega}\times \\  
&&
\left[1+\left(\frac{E}{E_{b1}}\right)^{1/\omega_1}\right]^{-(\alpha_2-\alpha_1)\omega_1}\left[1+\left(\frac{E}{E_{b2}}\right)^{1/\omega_2}\right]^{-(\alpha_3-\alpha_2)\omega_2}
\end{eqnarray}
which is a generalization of the one-break expression (see \cite{Lipari:2019jmk}).
In the above formula $E_0$ is an arbitrary reference energy that we fix to $1$ TeV, while $K$ gives the absolute normalization, and ${\alpha, \alpha_1, \alpha_2, \alpha_3, E_{\rm b}, E_{\rm b1}, E_{\rm b2}, \omega, \omega_1, \omega_2}$ are free parameters that determine the spectral shape.
The best-fit values of the parameters are the following: $K=4.37\times 10^{-5}\, \rm GeV^{-1}\, m^{-2}\, s^{-1}\, sr^{-1}$, $\alpha=2.95$, $\alpha_1=5.00$, $\alpha_2=-0.78$, $\alpha_3=4.01$, $E_{\rm b}=15.97$ TeV, $E_{\rm b1}=99.99$ TeV, $E_{\rm b2}=800.00$ TeV, $\omega=0.789$, $\omega_1=1.747$, $\omega_2=1.389$.
The result of our fit is shown with a blue line in Fig.~\ref{fig:KASKADEfit} in comparison with the proton fit provided by \cite{Dembinski:2017} with a red line.

\section{Comparison with the newly release WCDA-KM2A data}
\label{App: WCDA}
\begin{figure*}
\begin{center}
\subfigure[]{\includegraphics[width=0.45\textwidth]{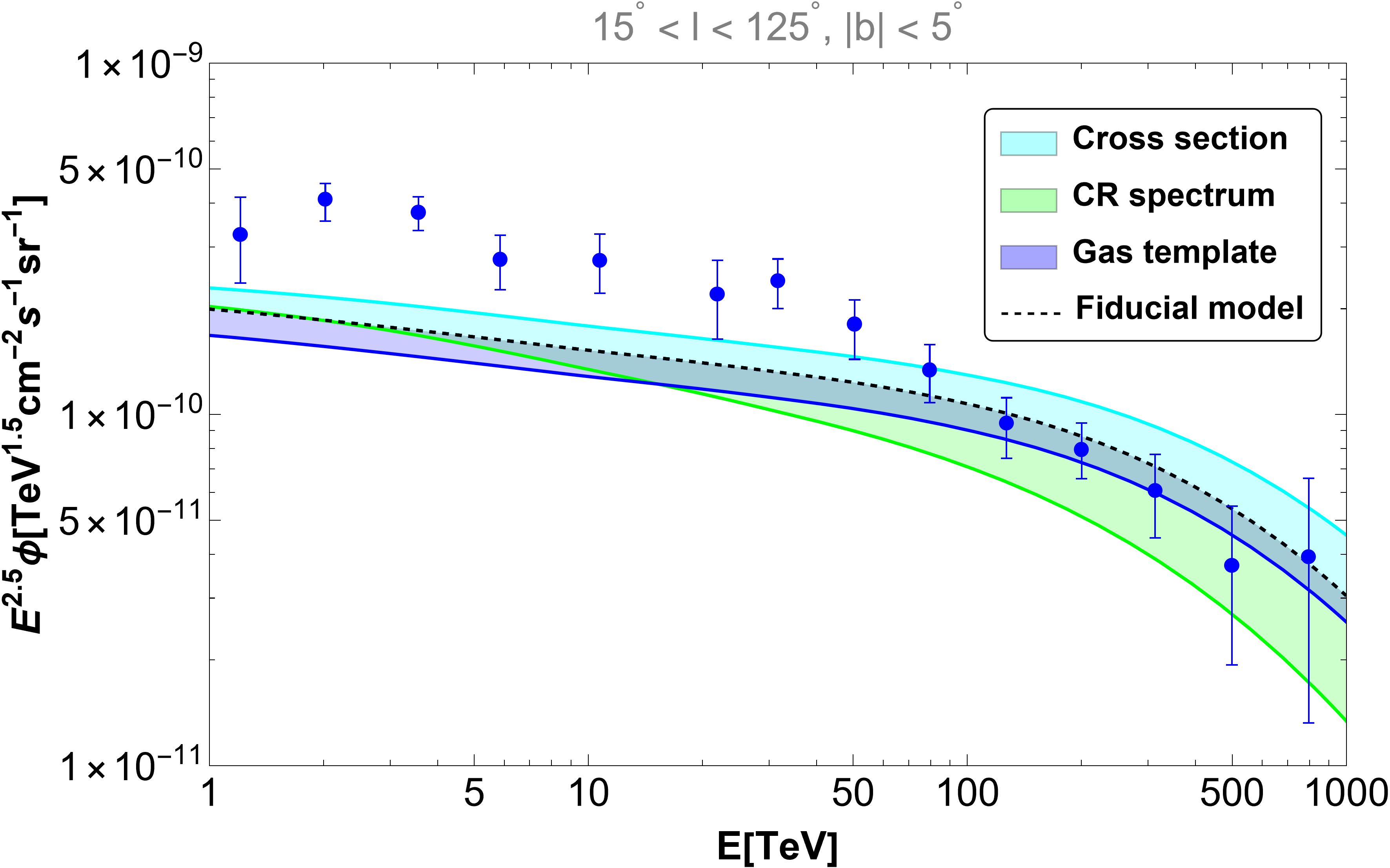}}
\subfigure[]{\includegraphics[width=0.45\textwidth]{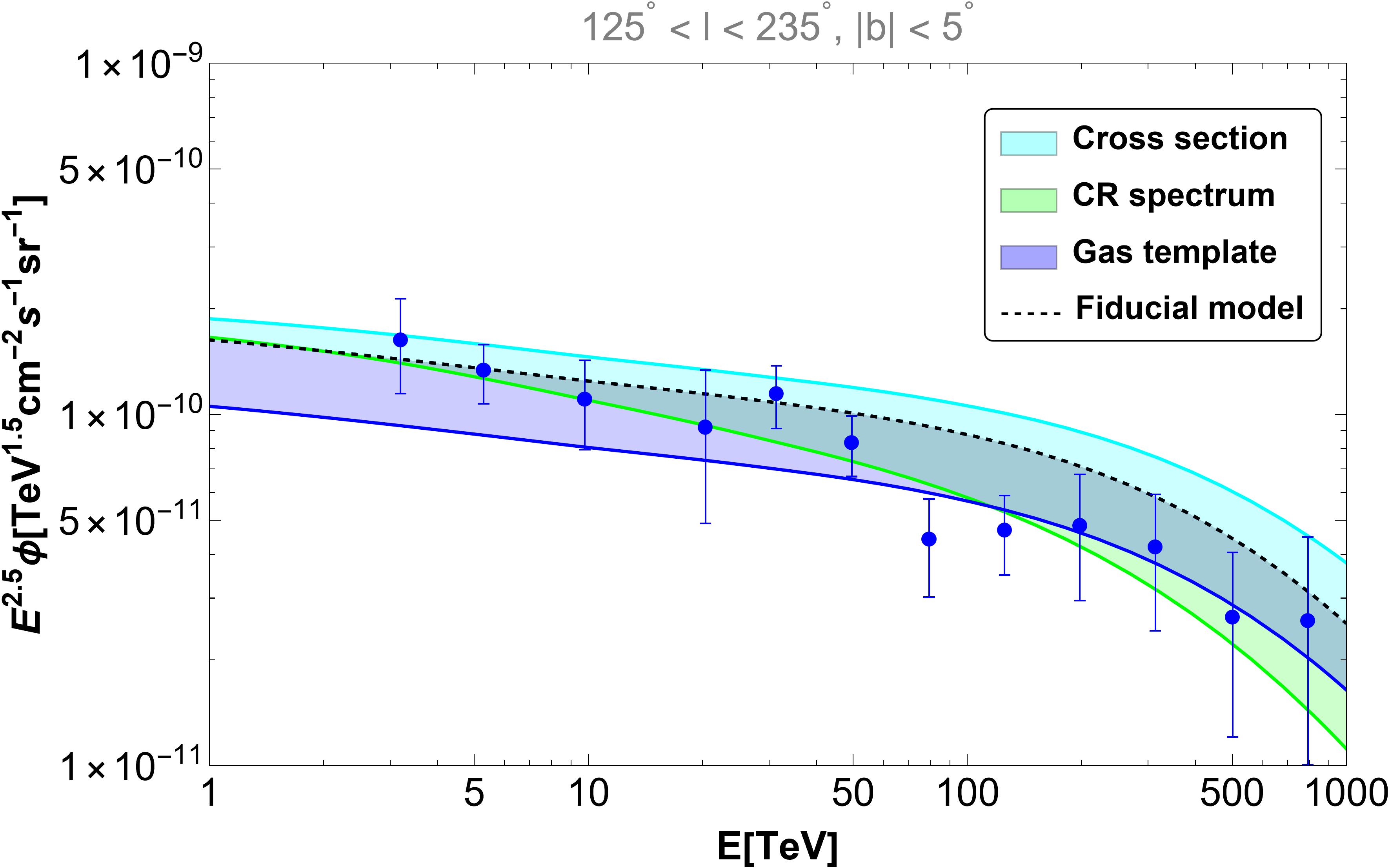}}
\caption{\small\em Same as Fig.~\ref{fig:LHAASOstandard} but with the blue data points representing the new WCDA-KM2A diffuse emission measurements provided by LHAASO \cite{LHAASO:2024lnz}.
}
\label{fig:LHAASOnew}
\end{center}
\end{figure*}

Here, we present a comparison between our model and the newly released LHAASO diffuse emission, obtained from the combined analysis of data from the WCDA and KM2A detectors.
As shown in Fig.~\ref{fig:LHAASOnew}, the main conclusion of this paper remains unchanged.
We do not add our unresolved sources on top of the diffuse emission model, as they are computed using the KM2A sensitivity threshold and the mask adopted in \cite{LHAASO:2023diff}.

\section{Comparison with the Fermi-LAT data}
\label{App:Fermi}
\begin{figure}[h!]
\begin{center}
\subfigure[]{\includegraphics[width=0.45\textwidth]{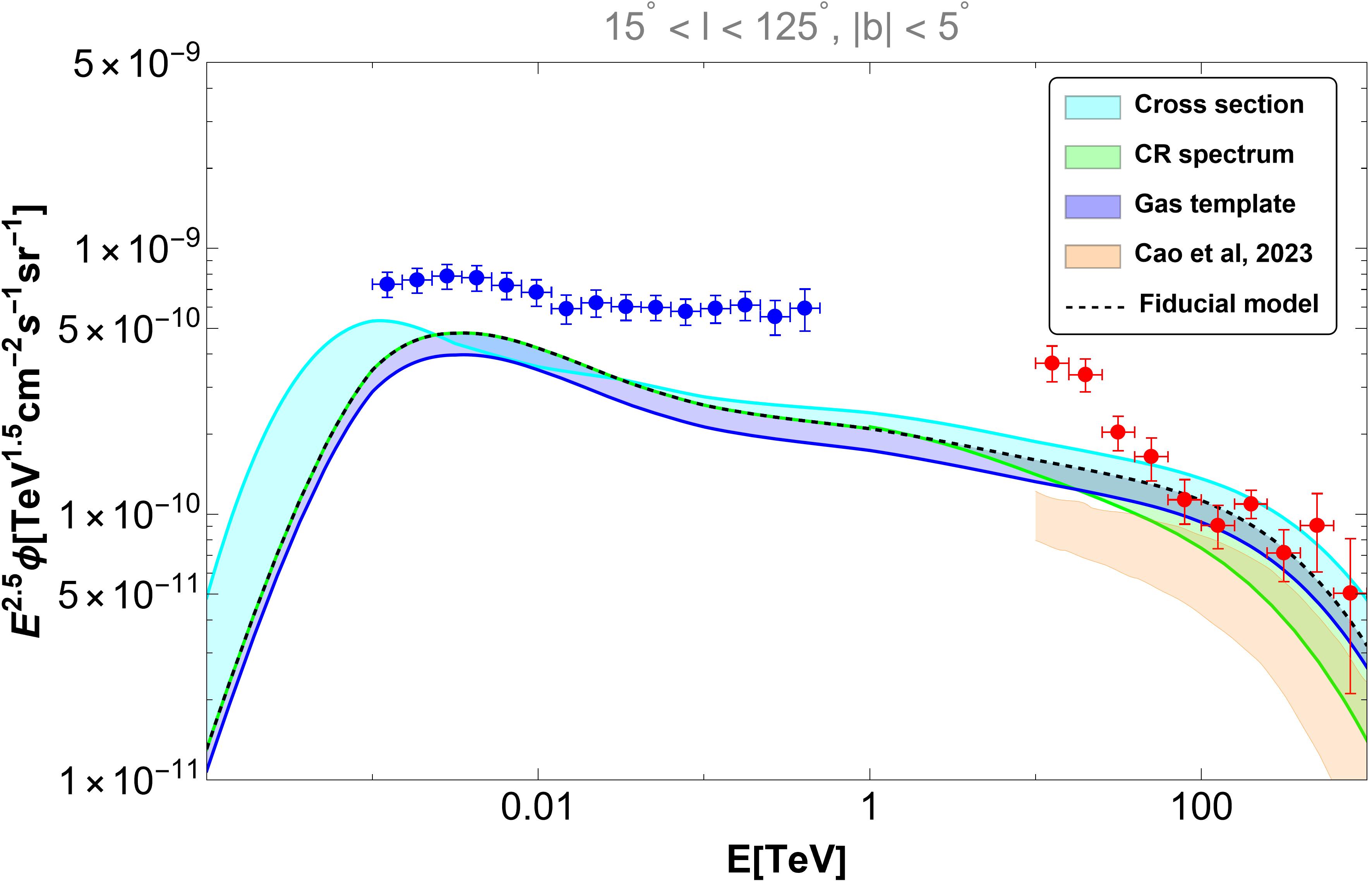}}
\subfigure[]{\includegraphics[width=0.45\textwidth]{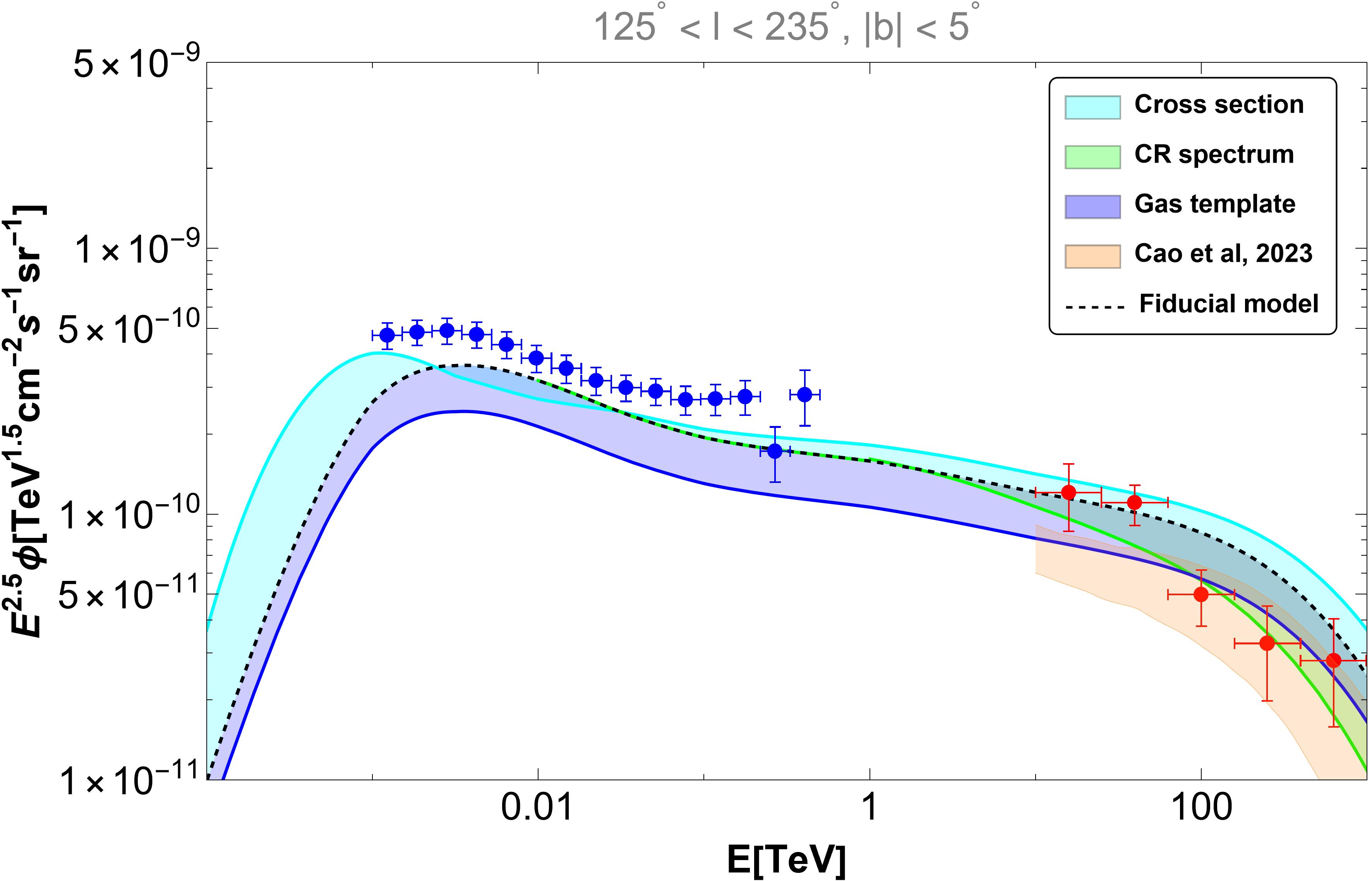}}
\caption{\small\em  Differential energy spectra of diffuse gamma-rays from the Galactic plane in the two angular regions probed by the LHAASO detector. Red data points are the measurements provided by LHAASO-KM2A \citep{LHAASO:2023diff} and blue data points are the Fermi-LAT data provided in \citep{Zhang:2023ajh}.}
\label{Fig: fermi}
\end{center}
\end{figure}

In Fig.~\ref{Fig: fermi}, we compare our hadronic diffuse emission model with data over an extended energy range, from $1$ GeV to $1$ PeV, using, in the sub-TeV range, the Fermi-LAT data provided by \cite{Zhang:2023ajh}, which were obtained through the same masking procedure applied to the LHAASO-KM2A measurements \cite{LHAASO:2023diff}.

It can be noticed that the hadronic diffuse emission underestimates the Fermi-LAT data in the GeV energy range.
This discrepancy is not unexpected and partially arises from the fact that our model does not include inverse Compton and bremsstrahlung components, which provide a non-negligible contribution to the diffuse emission at $few$ GeV.
Another missing component is the contribution from unresolved sources in the Fermi-LAT energy range.
Our unresolved source model is tuned to high-energy data and specifically calculated for the LHAASO KM2A detector and therefore cannot be used in the GeV energy range.

\end{document}